\documentclass[12pt]{article}
\usepackage{epsfig,graphicx}
\usepackage{amsmath}
\usepackage{amssymb}
\usepackage{array}
\newcommand{\beq}{\begin{equation}}
\newcommand{\eeq}{\end{equation}}
\newcommand{\bea}{\begin{eqnarray}}
\newcommand{\eea}{\end{eqnarray}}

\newcommand{\OMIT}[1]{{}}

\newcommand\spur{\raise.15ex\hbox{/}\kern-.57em }

\newcommand{\CL}{\mathcal{L}}

\newcommand{\upmns}{{\hat U}}
\newcommand{\upmnsdag}{{\hat U^\dagger}}
\newcommand{\upmnstar}{{\hat U^*}}
\newcommand{\XLFV}{\Delta}
\newcommand{\LLN}{{\Lambda_{\rm LN}}}
\newcommand{\LLNsq}{{\Lambda^2_{\rm LN}}}
\newcommand{\LLFV}{{\Lambda_{\rm LFV}}}
\newcommand{\LLFVsq}{{\Lambda^2_{\rm LFV}}}
\newcommand{\LLFVfourth}{{\Lambda^4_{\rm LFV}}}
\newcommand{\hc}{\text{h.c.}}
\newcommand{\lsim}{
\mathrel{\hbox{\rlap{\hbox{\lower4pt\hbox{$\sim$}}}\hbox{$<$}}}}
\newcommand{\gsim}{
\mathrel{\hbox{\rlap{\hbox{\lower4pt\hbox{$\sim$}}}\hbox{$>$}}}}
\def\npb#1#2#3{    {Nucl. Phys.}~B {\bf #1}, #3 (#2)}
\def\plb#1#2#3{    {Phys. Lett.}~B {\bf #1}, #3 (#2)}

\begin{document}

\begin{flushright}
KRL-MAP-316\\
UCSD/PTH-06-01\\
January 2006
\end{flushright}
\vspace{1.0 true cm}
\begin{center}
{\Large {\bf  Phenomenology of Minimal Lepton Flavor Violation}}\\
\vspace{1.0 true cm}
{\large Vincenzo Cirigliano${}^a$ and Benjam\'\i{}n Grinstein${}^b$} \\
\vspace{0.5 true cm}
${}^a$ {\sl California Institute of Technology, Pasadena, CA 91125} \\
\vspace{0.2 true cm}
${}^b$ {\sl Department of Physics, University of California at San Diego, 
La Jolla, CA 92093} \\
\vspace{0.2 true cm}

\end{center}
\vspace{0.5cm}

\begin{abstract}

We extend the effective theory of Minimal Lepton Flavor Violation
(MLFV) by including four-lepton operators.  We compute the rates for
$\mu \to 3 e$ and $\tau \to 3 {\ell}$ decays and point out
several new ways to test the hypothesis of MLFV. 
We also investigate to what extent it will be possible from (future)
experimental information to pin down the contributions of different
effective operators.  
In particular we look for experimental handles on quark-lepton
operators of the type $\bar\ell_i \Gamma \ell_j \times \bar q \Gamma
q$ by working out their contribution to hadronic processes such as
$\tau \to \mu \pi^0$, $\pi^0 \to \mu \bar{e}$, $\Upsilon \to \tau
\bar{\mu}$, as well as to purely leptonic decays such as $\mu \to 3 e$
through loop effects.
\end{abstract}

\section{Introduction}

In a recent work~\cite{Cirigliano:2005ck} we have extended the notion
of Minimal Flavor Violation (MFV)~\cite{Georgi,Hall:1990ac,MFV} to the
lepton sector of beyond the Standard Model (SM) theories.  The MFV
hypothesis states that the irreducible sources of (lepton) flavor
symmetry breaking are linked in a minimal way to the structures
generating the observed pattern of fermion masses and mixing. While
this idea has a straightforward and unique realization in the quark
sector~\cite{Georgi,Hall:1990ac,MFV} (the SM Yukawas are the only sources of
quark-flavor symmetry breaking), the situation in the lepton sector is
different, mainly due to the possibility/necessity to break the $U(1)$
symmetry associated with total lepton number.

The MFV framework has two particularly attractive features. On one hand,
it implies a suppression of FCNC processes induced by new degrees of
freedom at the TeV scale to a level which is consistent with present
experimental constraints. Moreover, it provides a predictive
framework that links the FCNC couplings to the fermion spectrum and
mixing structure (up to an overall scale factor). For leptons it
relates lepton-flavor mixing in the neutrino sector to lepton-flavor
violation in the charged lepton sector.

In ref.~\cite{Cirigliano:2005ck} we found two ways to define the
sources of flavor symmetry breaking in the lepton sector in a minimal
and thus very predictive way: i) a scenario where the left-handed
Majorana mass matrix is the only irreducible source of flavor symmetry
breaking; and ii) a scenario with heavy right-handed neutrinos, where
the Yukawa couplings define the irreducible sources of flavor symmetry
breaking and the right-handed Majorana mass matrix has a trivial
flavor structure. 

For each scenario, according to the MLFV symmetry principle, we have
constructed the basis of dimension six operators contributing to
processes with only two charged leptons~\cite{Cirigliano:2005ck}.  In
this context we have investigated the sensitivity of
processes such as $\ell_i \to \ell_j \gamma$ and $\mu \to e$
conversion to the scales of lepton flavor ($\LLFV$) and lepton number
($\LLN$) violation and we have pointed out distinctive predictions of 
MLFV  as a function of $s_{13}$ and the CP violating
phase $\delta$. 

In this paper we extend previous work in several respects:
\begin{itemize}
\item we enlarge the MLFV effective theory to include operators
contributing to four-lepton processes and we give predictions
for decays such as $\mu \to 3 e$ and $\tau \to 3 {\ell}$.
\item we investigate to what extent it will be possible to reconstruct
from (future) experimental information the dynamics of a given MLFV
model, which amounts to pin down the relative size of the couplings
appearing in the effective lagrangian.
\end{itemize}

We tackle this second issue by working out a number of predictions for
LFV decay rates. In particular, we study how LFV decays involving
hadrons can be used to probe operators involving two leptons and two
quarks (such as $ \bar\ell_i \Gamma \ell_j \times \bar q \Gamma q$).
We then show how, under suitable assumptions, the decay $\mu \to 3 e$
can also be used to probe operators of the type $\bar\ell_i \Gamma
\ell_j \times \bar q \Gamma q$ through loop effects.

The paper is organized as follows. In section~\ref{sect:operators} we
recall the two realizations of MLFV introduced in
Ref.~\cite{Cirigliano:2005ck} and give the complete basis of dimension
six effective operators, including four-lepton operators.  In
section~\ref{sect:mu3etree} we present results for the rates of
various $\ell\to\ell'\ell''\bar\ell'''$ decays at tree level in the
effective field theory, explore their phenomenology, and point out new
testable predictions of the MLFV scenario.  In
section~\ref{sect:hadronic} we discuss LFV decays involving hadrons,
while in section~\ref{sec:no4L} we calculate the loop-induced
contribution to $\mu \to 3e$ from four-fermion operators involving two
quark fields. 
We present our conclusions in section~\ref{sect:conclusions}.  
Details concerning the phase space integration, the renormalization
group analysis of section~\ref{sec:no4L} and the matching to chiral
perturbation theory are given respectively in
appendices~\ref{sect:app1},\ref{app-RG}, and \ref{app:chpt}. 

\section{Extended Operator Basis}
\label{sect:operators}
 
In this section we recall the two scenarios of MLFV identified in
Ref.~\cite{Cirigliano:2005ck} and present the dimension six operator
basis, extended to cover four-lepton processes.  In order to formulate
the minimal flavor violation hypothesis for leptons, one needs to
identify the field content of the theory, the flavor symmetry group
and the irreducible sources of symmetry breaking.  Two possibilities
arise (see ref.~\cite{Cirigliano:2005ck} for details).
%

\underline{\it Minimal field content}: In this scenario the flavor
symmetry group is $G_{\rm LF} = SU(3)_L\times SU(3)_E $, acting on
three left-handed lepton doublets $L_{L}^i$ and three right-handed
charged lepton singlets $e_{R}^i$ (SM field content).  The breaking of
the $U(1)_{\rm LN}$ is independent from the breaking of $G_{\rm LF}$ and
is associated to a very high scale $\LLN$.  The irreducible sources of
lepton-flavor symmetry breaking are $\lambda_e^{ij}$ and $g_\nu^{ij}$,
defined by\footnote{~Throughout this paper we use four-component
spinor fields, and $\psi^c=-i\gamma^2\psi^*$ denotes the charge
conjugate of the field $\psi$. We also use $v = \langle H^0 \rangle
\simeq 174$ GeV.}
\begin{align}
\label{eq:minlag}
\CL_{\text{Sym.Br.}} &= - \lambda_e^{ij} \,\bar e^i_R(H^\dagger L^j_L)
-\frac1{2 \LLN}\,g_\nu^{ij}(\bar L^{ci}_L\tau_2 H)(H^T\tau_2L^j_L)+\hc
\end{align}  

\underline{\it Extended field content}: In this scenario the maximal flavor
group is $G_{\rm LF} \times SU(3)_{\nu_R}$, acting on three
right-handed neutrinos, $\nu^i_R$, in addition to the SM fields.  The
right-handed neutrino mass term is flavor diagonal ($M_\nu^{ij}=M_\nu
\delta^{ij}$, $|M_\nu|\gg v$) and its effect is to break $U(1)_{\rm
LN}$ as well as to break $SU(3)_{\nu_R}$ to $O(3)_{\nu_R}$.  The
remaining lepton-flavor symmetry is broken only by two irreducible
sources, $\lambda_e^{ij}$ and $\lambda_\nu^{ij}$, defined by 
\begin{align}
\label{eq:extlag}
\CL_{\text{Sym.Br.}}  & = - \lambda_e^{ij} \,\bar e^i_R(H^\dagger
L^j_L) +i \lambda_\nu^{ij}\bar\nu_R^i(H^T \tau_2L^j_L)+\hc  
\end{align}

In most SM extensions, at some scale $\LLFV$ above the electroweak
scale and well below $\Lambda_{\rm LN}$ (or $M_\nu$) there are new
degrees of freedom carrying lepton flavor quantum numbers.  As long as
the underlying model respects MLFV, at scales below $\LLFV$ the
physics of lepton flavor violation is described by the following
effective lagrangian (up to higher dimensional operators)
\beq
\label{eq:eff-lag}
\CL=\frac{1}{\LLFVsq}
\sum_{i=1}^5 \Big(c_{LL}^{(i)} 
O_{LL}^{(i)} +c_{4L}^{(i)} 
O_{4L}^{(i)}\Big)+   \frac{1}{\LLFVsq} \left(
\sum_{j=1}^2 c_{RL}^{(j)} O_{RL}^{(j)}  +\hc  \right) \ , 
\eeq
with operators defined by:
\begin{equation}
\label{eq:ops-listed}
\begin{aligned} 
O^{(1)}_{LL} &= \bar L_L\gamma^\mu \XLFV L_L\; H^\dagger iD_\mu H\\
O^{(2)}_{LL} &= \bar L_L\gamma^\mu\tau^a \XLFV L_L\; H^\dagger \tau^aiD_\mu H\\
O^{(3)}_{LL} &= \bar L_L\gamma^\mu \XLFV L_L\; \bar Q_L\gamma_\mu Q_L \\
O^{(4d)}_{LL} &= \bar L_L\gamma^\mu \XLFV L_L\; \bar d_R\gamma_\mu d_R \\
O^{(4u)}_{LL} &= \bar L_L\gamma^\mu \XLFV L_L\; \bar u_R\gamma_\mu u_R \\
O^{(5)}_{LL} &= \bar L_L\gamma^\mu\tau^a \XLFV L_L\;\bar Q_L\gamma_\mu
\tau^a Q_L 
\end{aligned} 
\qquad
\begin{aligned} 
O^{(1)}_{RL} & = g^\prime 
 H^\dagger\bar e_R\sigma^{\mu\nu} \lambda_e^{\phantom{\dagger}}
\XLFV L_L\, B_{\mu\nu}\\
O^{(2)}_{RL} & = g 
 H^\dagger\bar e_R\sigma^{\mu\nu}\tau^a \lambda_e^{\phantom{\dagger}}\XLFV L_L\, W^a_{\mu\nu} \\
O^{(3)}_{RL} & = (D_\mu H)^\dagger \bar e_R \lambda_e^{\phantom{\dagger}}\XLFV D_\mu L_L\\
O^{(4)}_{RL} & =   
 \bar e_R  \lambda_e^{\phantom{\dagger}}\XLFV L_L\, \bar Q_L \lambda_D d_R \\ 
O^{(5)}_{RL} & =  
 \bar e_R \sigma^{\mu\nu} \lambda_e^{\phantom{\dagger}}\XLFV L_L\, 
 \bar Q_L \sigma_{\mu\nu} \lambda_D d_R \\
O^{(6)}_{RL} & =   
 \bar 
e_R  \lambda_e^{\phantom{\dagger}}\XLFV L_L\, \bar u_R  \lambda_U^\dagger  i\tau^2 Q_L \\ 
O^{(7)}_{RL} & =  
 \bar e_R \sigma^{\mu\nu} \lambda_e^{\phantom{\dagger}}\XLFV L_L\, 
 \bar u_R \sigma_{\mu\nu}  \lambda_U^\dagger   i\tau^2 Q_L
\end{aligned} 
\end{equation}  
and
\begin{equation}
\label{eq:4Lops}
\begin{aligned} 
O^{(1)}_{4L} & =\bar L_L\gamma^\mu \XLFV L_L\; \bar L_L\gamma_\mu L_L \\
O^{(2)}_{4L} & =\bar L_L\gamma^\mu\tau^a \XLFV L_L\;\bar L_L\gamma_\mu\tau^a L_L\\
O^{(3)}_{4L} & =\bar L_L\gamma^\mu \XLFV L_L\; \bar e_R\gamma_\mu e_R \\
O^{(4)}_{4L} & =\delta_{nj}\delta^*_{mi}\bar L^i_L \gamma^\mu  L^j_L  ~\bar L^m_L \gamma^\mu  L^n_L \\
O^{(5)}_{4L} & =\delta_{nj}\delta^*_{mi}\bar L^i_L \gamma^\mu \tau^a L^j_L  ~\bar L^m_L 
\gamma^\mu \tau^a  L^n_L  \ . 
\end{aligned} 
\end{equation}  
The MLFV hypothesis fixes the couplings $\Delta$ and $\delta$ to be:
\beq
\label{eq:XLFV}
\XLFV =  \begin{cases}
 \frac{\LLNsq}{v^4}  ~ \upmns  m^2_\nu  \upmnsdag~ &\text{minimal field  content}\\
 \frac{M_\nu}{v^2}  ~ \upmns  m_\nu  \upmnsdag~ &\text{extended field
  content, CP limit}
\end{cases}
\eeq
and
\beq
\delta=\delta^T= \begin{cases}
\frac{\LLN}{v^2} \, \upmnstar  m_\nu  \upmnsdag~ &\text{minimal field
  content}\\
\frac{M_\nu}{v^2} \, \upmnstar  m_\nu   \upmnsdag~ &\text{extended field
  content}
\end{cases}
\eeq
in terms of the diagonal light neutrino mass matrix $m_\nu$ and the
PMNS matrix $\hat U$.  To arrive at the above results we have used   
that $\lambda_e \ll 1$ and assumed perturbative behavior in $\Delta$
and $\delta$, stopping at leading order in $(\LLN m_\nu)/v^2$ (and
$(M_\nu m_\nu)/v^2$).  Therefore, for consistency, in the extended
field case the operators $O_{4L}^{(4),(5)}$ should be dropped.

The explicit form of the couplings $\Delta_{ij}$ in terms of neutrino
mass splitting and mixing angles (using the parameterization of the
PMNS matrix $\hat U$ of Ref.~\cite{PDG}) is reported in
Ref.~\cite{Cirigliano:2005ck} for both minimal and extended field
scenarios.  
Indicating by $s$ and $c$ the sine and cosine of the solar mixing
angle and by $s_{13} \equiv \sin \theta_{13}$~\cite{PDG}, the explicit
form of $\delta_{ij}$ in the case of minimal field content (the only
case we need) to first order in $s_{13}$ is,
\beq
\label{eq:deltas}
\begin{aligned}
\delta_{ee}&=\frac{\LLN}{v^2}\left[c^2e^{-i\alpha_1}m_{\nu_1}+
s^2  e^{-i\alpha_2}m_{\nu_2}
\right]\equiv\frac{\LLN}{v}\;d_{ee}\\
\delta_{e\mu}&=\frac{\LLN}{v^2}\frac1{\sqrt{2}}
\left[-sce^{-i\alpha_1}m_{\nu_1}+sc e^{-i\alpha_2}m_{\nu_2}
  +s_{13}e^{i\delta}m_{\nu_3}\right]\equiv\frac{\LLN}{v}\;d_{e\mu}\\
\delta_{\mu\mu}&=\frac{\LLN}{v^2}\frac12\left[s^2e^{-i\alpha_1}m_{\nu_1}+c^2  e^{-i\alpha_2}m_{\nu_2}
  +m_{\nu_3}\right]\equiv\frac{\LLN}{v}\;d_{\mu\mu}\\
\delta_{e\tau}&=\frac{\LLN}{v^2}\frac1{\sqrt{2}}
\left[sce^{-i\alpha_1}m_{\nu_1}-sc e^{-i\alpha_2}m_{\nu_2}
  +s_{13}e^{i\delta}m_{\nu_3}\right]\equiv\frac{\LLN}{v}\;d_{e\tau}\\
\delta_{\mu\tau}&=\frac{\LLN}{v^2}\frac12\left[-s^2e^{-i\alpha_1}m_{\nu_1}-c^2  e^{-i\alpha_2}m_{\nu_2}
  +m_{\nu_3}\right]\equiv\frac{\LLN}{v}\;d_{\mu\tau} \ . 
\end{aligned}
\eeq
Throughout this paper we will assume maximal atmospheric mixing and
the following central values for the remaining neutrino mixing
parameters~\cite{strumia}: $\Delta m^2_{\rm sol} = 8.0 \times 10^{-5}
\, {\rm eV}^2$, $\Delta m^2_{\rm atm} = 2.5 \times 10^{-3} \, {\rm
eV}^2$, $\theta_{\rm sol} = 33^o$.

\section{$\ell\to\ell'\ell''\bar\ell'''$ at tree level in the effective 
theory}
\label{sect:mu3etree}

Within the effective theory described in the previous section we can now
study decays of the type $\ell\to\ell'\ell''\bar\ell'''$.  At tree
level these decays receive contributions from the transition magnetic
moment operators $O^{(1),(2)}_{RL}$ (photon exchange), the four-lepton
operators $O^{(i)}_{4L}$, and from $O_{LL}^{(1),(2)}$ ($Z$ exchange).

\subsection{Rates}

The integrated decay rate for $\mu\to ee\bar e$ averaged over initial
polarizations and summed over final polarizations is:
\begin{equation}
\label{eq:rate1}
\Gamma_{\mu \to 3e} = \Gamma_{\mu \to e \nu \bar{\nu}}  \ 
\frac{v^4 |\Delta_{e \mu}|^2 }{\Lambda_{\rm LFV}^4} \  
\Big[ |a_+|^2+2|a_-|^2-8\text{Re}(a_0^*a_-)
-4\text{Re}(a_0^*a_+)+6I|a_0|^2 \Big]
\end{equation}
where $\Gamma_{\mu \to e \nu \bar{\nu}}= 
m_\mu^5/(v^4 \, 1536 \, \pi^3 )$, the  $a_{i}$ coefficients are  
\begin{equation}
\label{eq:ai}
\begin{aligned} 
a_+&=\sin^2\theta_w(c^{(1)}_{LL}+c^{(2)}_{LL})+c^{(3)}_{4L}\\
a_-&=(\sin^2\theta_w-\frac12)(c^{(1)}_{LL}+c^{(2)}_{LL})+
  c^{(1)}_{4L}+c^{(2)}_{4L} + \frac{2\delta_{e\mu}^{\phantom{*}}
\delta_{ee}^*}{\Delta_{e \mu}} \,    (c^{(4)}_{4L}+c^{(5)}_{4L})\\
a_0&=2e^2(c^{(1)}_{RL}-c^{(2)}_{RL})^*
\end{aligned} 
\end{equation}  
and $I$ is a phase space integral given by (see appendix~\ref{sect:app1} for 
details)
\begin{equation}
\label{eq:Ips}
I\approx\frac83\ln\left(\frac{m_\mu}{m_e}\right)
-\frac43 \ln(2)-\frac{26}{9}  +{\cal O}\left(\frac{m_e}{m_\mu}\right) ~ . 
\end{equation}
Note that $a_{-}$ depends not only on the Wilson coefficients but also
on the ratio of effective FCNC couplings mediating $\mu \to e$
transitions, namely $\delta_{e\mu}^{\phantom{*}}\delta_{ee}^*$ and
$\Delta_{e \mu}$. We shall study the consequences of this in Section 
\ref{sect:pheno2}. 

The rate for $\tau\to\mu\mu\bar\mu$ and for $\tau\to ee\bar e$ is 
given similarly, replacing $\Delta_{e\mu}\to \Delta_{\mu\tau}$,
$\delta_{e\mu}^{\phantom{*}}\delta_{ee}^*\to
\delta_{\mu\tau}^{\phantom{*}}\delta_{\mu\mu}^*$ and $\Delta_{e\mu}\to
\Delta_{e\tau}$, $\delta_{e\mu}^{\phantom{*}}\delta_{ee}^*\to
\delta_{e\tau}^{\phantom{*}}\delta_{ee}^*$, respectively, and
$m_\mu\to m_\tau $. 

The integrated rate for $\tau^-\to e^-\mu^-\mu^+$ is given by
\beq
\label{eq:rate2}
\Gamma_{\tau \to e \mu \bar\mu} = 
\Gamma_{\tau \to e \nu \bar{\nu}}  \ 
\frac{v^4 |\Delta_{e \tau}|^2 }{\Lambda_{\rm LFV}^4} \  
\left[|a_+|^2+|\tilde{a}_-|^2-4\text{Re}[a_0^*(a_++\tilde{a}_-)]+
12 \tilde I|a_0|^2\right]
\eeq
where $a_+, a_0$ are given in Eq.~(\ref{eq:ai}) while  
\begin{equation}
\tilde{a}_- =(\sin^2\theta_w-\frac12)(c^{(1)}_{LL}+c^{(2)}_{LL})+
  c^{(1)}_{4L}+c^{(2)}_{4L} + \frac{4\delta_{\mu \tau}^{\phantom{*}}
\delta_{\mu e}^*}{\Delta_{e \tau}} \,    (c^{(4)}_{4L}+c^{(5)}_{4L}) ~ , 
\end{equation}  
and $\tilde I$ is a phase space integral reported in
appendix~\ref{sect:app1}.  The rate for $\tau^-\to \mu^-e^+e^-$ is
obtained by exchanging the labels $e\leftrightarrow \mu$ in $\Delta$
and $\delta$. 

Finally, the rate for $\tau^-\to\mu^-\mu^-e^+$ is given by
Eq.~(\ref{eq:rate1}) replacing $\mu \to \tau$, but 
with
\beq
\begin{aligned} 
a_+&=a_0=0\\
a_-&=\frac{2\delta_{e\tau}^{\phantom{*}}\delta_{\mu\mu}^*}{ 
\Delta_{e \tau}
}(c^{(4)}_{4L}+c^{(5)}_{4L})
\end{aligned} 
\end{equation} 
(no photon and $Z$ exchange contribution) and similarly for $\tau^-\to
e^-e^-\mu^+$, exchanging the labels $e\leftrightarrow \mu$ in
$\delta$.

In order to explore the phenomenology of these results, it is
convenient to express the rates in terms of dimensionless couplings
with the large scale $\LLN$ (for minimal field content) or $M_\nu$
(for extended field content) factored out. Hence we introduce the
quantities $a_{ij}$, $b_{ij}$ and $d_{ij}$ defined through
\begin{equation}
\Delta_{ij}=\begin{cases}\frac{\LLNsq}{v^2}a_{ij}& 
\text{minimal field content}\\
\frac{M_\nu}{v}b_{ij}& \text{extended field content}\end{cases}
\end{equation}
and Eq.~(\ref{eq:deltas}). These quantities depend only on low energy
masses and mixing angles, and can be readily estimated.  As an example
we report the resulting expression for the $\mu \to 3 e$ rate:
\begin{equation}
\begin{aligned} 
\Gamma_{\mu \to 3e}/ \Gamma_{\mu \to e \nu \bar{\nu}}  &=  
\Big[ |a_+|^2+2|a_-|^2-8\text{Re}(a_0^*a_-)
-4\text{Re}(a_0^*a_+)+6I|a_0|^2 \Big] \times
\\ 
& \times 
\begin{cases} \left(\frac{\LLN}{\LLFV}\right)^4 |a_{e \mu}|^2 
& \text{minimal field content}\\
\left(\frac{v M_\nu}{\LLFV^2}\right)^2 |b_{e \mu}|^2 
& \text{extended field content}\end{cases} ~ . 
\end{aligned} 
\label{eq:rate3}
\end{equation}

\subsection{Phenomenology: extended field content} 
\label{sect:pheno1}

Let us consider first the MLFV realization with extended field
content, where the discussion is somewhat simpler.  In this case the
contributions of $O^{(4)}_{4L}$ and $O^{(5)}_{4L}$ to $\mu\to eee$ are
negligible, since they involve the combination $\delta_{\mu
e}\delta^*_{ee}$ which is suppressed by a small neutrino mass relative
to the contributions to the amplitude that are proportional to
$\Delta_{\mu e}$. As a consequence, the transition between families
$i$ and $j$ is governed just by $b_{ij}$, the only allowed FCNC
effective coupling.  Moreover, for Wilson coefficients of $O(1)$, the
expression in square brackets in Eq.~(\ref{eq:rate3}) is $O(1)$ and
therefore, for a given value of $(v M_\nu/\LLFV^2)$, the branching
fraction for $\mu \to 3e$ is determined by $|b_{e\mu}|^2$.

\begin{figure}[t]
\centering
\epsfxsize=7.7cm
\epsffile{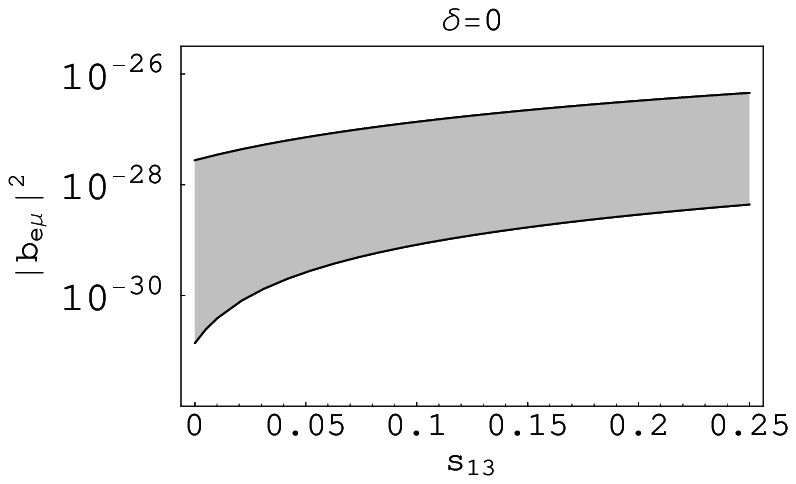}
\epsfxsize=7.7cm
\epsffile{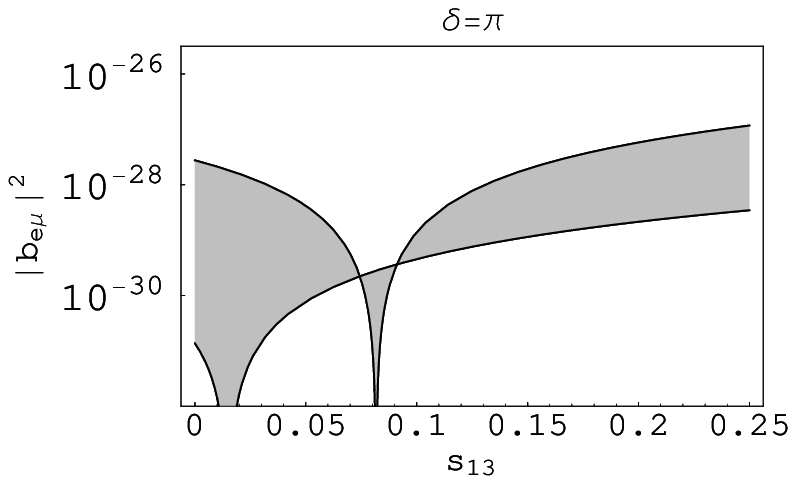}
\caption{\label{fig:b-plot} $|b_{e\mu}|^2$ as a function of $s_{13}$
for normal neutrino mass hierarchy and the PMNS phase $\delta=0$
(left) or $\delta=\pi$ (right).  The shaded bands correspond to
variations of the lightest neutrino mass in the range $0 \leq
m_{\rm min} \leq 0.2 \ {\rm eV}$.  The upper edge of the
bands corresponds to $m_{\rm min} = 0$.  }
\end{figure}

Fig.~\ref{fig:b-plot} shows the behavior of $|b_{e\mu}|^2$ as a
function of (the poorly determined mixing angle) $s_{13}$ varying the
lightest neutrino mass in the experimentally allowed range $0 \leq
m_{\rm min} \leq 0.2 \ {\rm eV}$~\cite{PDG}, in the case of
normal ordering of the neutrino mass spectrum.  The plots correspond
to the two CP conserving values of the phase $\delta$ in the PMNS
matrix: $\delta = 0$ (left panel) and $\delta = \pi$ (right panel).
As can be seen, the present uncertainties in $s_{13}$ and the absolute
scale of neutrino spectrum induce a variation of $|b_{e\mu}|^2$ over a
couple of orders of magnitude.  Similar results apply in the case of
inverted spectrum.  For $M_\nu=6\times10^7\LLFVsq/v$, which saturates
the perturbative Yukawa coupling bound on $M_\nu$ when $\LLFV=50$~TeV,
one obtains $\mu \to 3e$ branching fractions of order $10^{-12}$,
comparable to the 90\% C.L. limit of
$1.0\times10^{-12}$~\cite{Bellgardt:1987du}.  The $\tau \to 3 \mu$
branching ratio is governed by $|b_{\mu \tau}|^2$, which does not
depend on $s_{13}$ and for given $m_{\rm min}$ is typically
two orders of magnitude larger than $|b_{e \mu}|^2$, leading to
$B_{\tau \to 3 \mu} \sim 10^{-10}$.  This is below current
experimental sensitivities~\cite{Aubert:2003pc,Yusa:2004gm}.

The inclusion of 4-lepton processes in the phenomenological analysis
offers more ways to test the hypothesis of minimal lepton flavor
violation.  We report below two noteworthy features, which with
appropriate changes (see next subsection) extend to the minimal field
scenario as well. 

\begin{itemize}
\item 
As pointed out in~\cite{Cirigliano:2005ck}, testable predictions of
MLFV involve ratios of FCNC transitions {\em between two different
families} (e.g.  $\mu \to e$ vs $\tau \to \mu$).  In the case of
three-lepton final states, several operators contribute and they 
pick a different phase space weight, in principle. So predictions
are not as clean as in the case of lepton-photon final state.  We find
\begin{equation}
\label{eq:ratio3lept}
\frac{B_{\mu \to 3e}}{B_{\tau \to 3\mu}} = 
\frac{ |b_{e\mu}|^2}{|b_{\mu \tau}|^2}  
\frac{I_{\mu \to 3 e}}{I_{\tau \to 3 \mu}} \ 
\displaystyle\frac{1 + \displaystyle\frac{r_c}{I_{\mu \to 3 e}}}{1 + 
\displaystyle\frac{r_c}{I_{\tau \to 3 \mu}}} \ , 
\end{equation}
where the phase space integrals are $I_{\mu \to 3e}=9.886$, $I_{\tau
\to 3 \mu}=3.26$ and $r_c$ is the following combination of 
Wilson coefficients (in the absence of contact operators $r_c=0$): 
\begin{equation}
\label{eq:rc}
r_c = \frac{1}{6 |a_0|^2}
\left( |a_+|^2 + 2 |a_-|^2 -8\text{Re}(a_0^*a_-)-
4\text{Re}(a_0^*a_+) \right)~. 
\end{equation}
Modulo the last ($r_c$ dependent) factor which is of $O(1)$,
Eq.~(\ref{eq:ratio3lept}) has the same structure of the prediction for
the ratio of radiative decays $B_{\mu \to e\gamma}/B_{\tau \to \mu
\gamma}= |b_{e\mu}|^2/|b_{\mu \tau}|^2$, and will offer another way to
test the MLFV pattern $B_{\tau \to \mu} \gg B_{\mu \to e} \sim B_{\tau
\to e}$.  A precise test requires knowledge of the ratio $r_c$, 
which can be determined if additional LFV modes are observed, 
as we now discuss.

\item 
Ratios of FCNC transitions {\em between the same two families} (e.g.
$\mu \to e \gamma$ vs $\mu \to 3 e$ or $\tau \to \mu \gamma$ vs $\tau
\to 3 \mu$ and $\tau \to \mu e \bar{e}$)  are determined by
known phase space factors and ratios of various Wilson coefficients.  
If both $\mu\to e\gamma $ and $\mu\to eee$ processes are observed one 
can begin to disentangle the effects of photon exchange from those of
contact interactions, through the ratio of rates 
\begin{equation}
\label{eq:ratio3lept2}
\frac{\Gamma_{\mu \to 3e}}{\Gamma_{\mu \to e\gamma}} =
\frac{\alpha}{4 \pi} \  I_{\mu \to 3 e} \times \Big[ 1 + 
\displaystyle\frac{r_c}{I_{\mu \to 3e}} \Big]~, 
\end{equation} 
with $I_{\mu \to 3e}$ defined in Eq.~(\ref{eq:Ips})
and $r_c$ in Eq.~(\ref{eq:rc}).  The above ratio  
reduces to $\alpha/4 \pi  \ I_{\mu \to 3 e}$ in the absence of 
contact interactions and offers a way to experimentally determine 
$r_c$.  
One obtains the same result for $\Gamma_{\tau \to 3\mu}/\Gamma_{\tau
\to \mu \gamma}$ (with $I_{\mu \to 3e} \to I_{\tau \to 3 \mu}$ and the 
same $r_c$) and a
similar one for $\Gamma_{\tau \to \mu e \bar{e}}/\Gamma_{\tau \to \mu
\gamma}$.

\end{itemize}

\subsection{Phenomenology: minimal field content}
\label{sect:pheno2}

In the case of minimal field content, the analysis of 4-lepton
processes is complicated by the contributions of $O_{4L}^{(4)}$ and
$O_{4L}^{(5)}$, implying that there are two FCNC effective couplings
that mediate $\mu \to 3 e$: $a_{\mu e}$ and $d_{\mu e} d_{ee}^*$ (see
Eqs.~(\ref{eq:rate1}) and (\ref{eq:ai})).  This naturally leads us to
consider two cases. 

\begin{figure}[t]
\centering
\epsfxsize=9cm
\epsffile{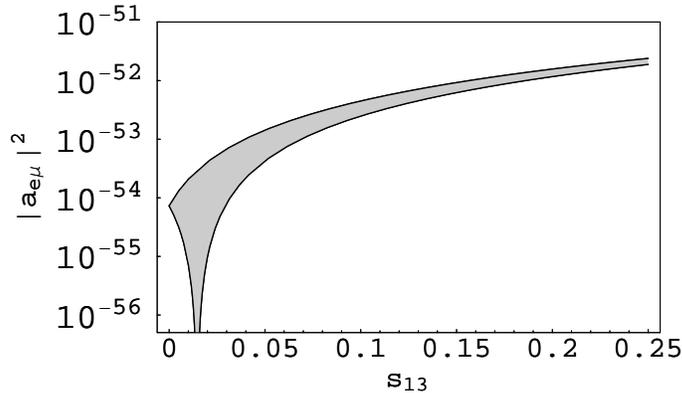}
\caption{\label{fig:a-plot} 
Range of $|a_{e\mu}|^2$ as a
function of $s_{13}$. The shaded band corresponds to varying the 
PMNS phase $\delta$ between $0$ and $2 \pi$.
The range of $|a_{e\tau}|^2$ is the same as this,
but is anti-correlated: the maximum of $|a_{e\mu}|^2$, achieved at
$\delta=0$, corresponds to the minimum of $|a_{e\tau}|^2$.   }
\end{figure}

(i) If the dynamics of the underlying MLFV model results in
$c^{(4)}_{4L}+c^{(5)}_{4L} \ll c_{4L}^{(i)}$ for $i \neq 4,5$, then the
analysis of $\mu \to 3 e$ processes parallels the one of the extended
field content given in the previous section, with the changes $|b_{\mu
e}|^2 \to |a_{\mu e}|^2$ and $(v M_\nu/\LLFV^2) \to (\LLN/\LLFV)^2$.
As before, for a given value of $\LLN/\LLFV$, $|a_{\mu e}|^2$
determines the $\mu \to 3e$ branching ratio.  We show in
Fig.~\ref{fig:a-plot} the range of allowed values of $|a_{e\mu}|^2$ as
a function of $s_{13}$, varying the PMNS phase $\delta$ between 0 and
$2 \pi$.  We see that, except for very small $s_{13}$, in the minimal
field content scenario a ratio $\LLN/ \LLFV=10^{10}$ implies a
branching fraction for $\mu\to ee\bar e$ as large as a few times
$10^{-12}$, comparable the 90\% C.L. limit of
$1.0\times10^{-12}$\cite{PDG}.  The range of
$|a_{e\tau}|^2$ is the same as for $|a_{e\mu}|^2$, but is
anti-correlated: the maximum of $|a_{e\mu}|^2$, achieved at $\delta=0$,
corresponds to the minimum of $|a_{e\tau}|^2$. The quantity
$a_{\mu\tau}$ is independent of $s_{13}$,
$|a_{\mu\tau}|^2=1.6\times10^{-51}$.  In this scenario
Eq.~(\ref{eq:ratio3lept}) remains valid with the substitution $|b_{i
j}|^2 \to |a_{i j}|^2$, and $r_c$ (given in Eq.~(\ref{eq:rc}))
involves just ratios of Wilson coefficients that can be determined
experimentally by considering the ratio $\Gamma_{\mu \to
3e}/\Gamma_{\mu \to e\gamma}$ as in Eq.~(\ref{eq:ratio3lept2}).

\begin{figure}[t]
\centering
\epsfxsize=7.7cm
\epsffile{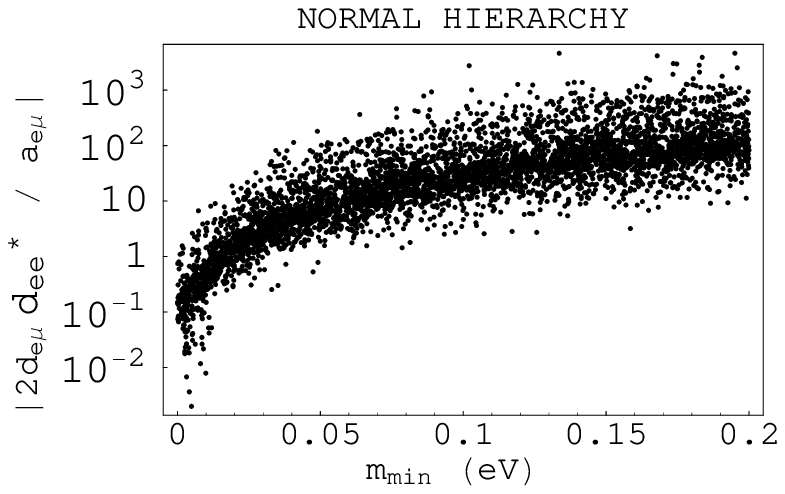}
\epsfxsize=7.7cm 
\epsffile{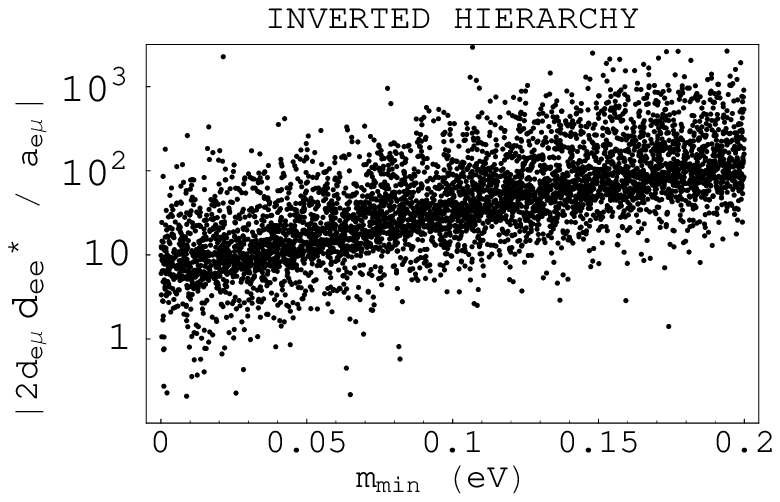}
\caption{\label{fig:dd-a-ratio} 
Range of the ratio
$|2d_{e\mu}d_{ee}^*/a_{e \mu}|$ as a function of $m_{\rm min} ({\rm
eV})$ in the case of normal hierarchy (left) and inverted hierarchy
(right).  We use the range $0 \leq s_{13} \leq 0.20$ and we allow the
CP violating phases in the PMNS matrix to vary between 0 and $2
\pi$. For a given value of $m_{\rm min}$, the upper edge of the range 
corresponds to $s_{13} \to 0$.}
\end{figure}

(ii) If the underlying dynamics is such that all Wilson coefficients
are of comparable magnitude, then it is important to address the
relative size of the FCNC couplings $a_{\mu e}$ and $d_{\mu e}
d_{ee}^*$, as there is a potential competition between them.  In
Fig.~\ref{fig:dd-a-ratio} we plot the ratio $r_{e \mu} \equiv
|2d_{e\mu}d_{ee}^*/a_{e \mu}|$ as a function of $m_{\rm min}$ for $0
\leq m_{\rm min} \leq 0.2$ eV, in the case of normal hierarchy (left)
and inverted hierarchy (right). The points correspond to a scanning in
parameter space with $0 \leq s_{13} \leq 0.20$ and $0 \leq
\alpha_{1,2}, \delta \leq 2 \pi$ for the CP violating phases in the
PMNS matrix.  For a given value of $m_{\rm min}$, the upper edge of
the range corresponds to $s_{13} \to 0$.  As can be seen, $r_{\mu e}$
displays a very strong dependence on the absolute mass scale of the
neutrino spectrum.  Typically we find $r_{e \mu}\gg 1$, except in the
normal hierarchy case and for $m_{\rm min} \to 0$.  Moreover, for a
given value of $m_{\rm min}$ there are regions of parameter space
where cancellations induced by CP violating phases 
drive $r_{\mu e}$ to zero.

In the general case (ii),
Eq.~(\ref{eq:ratio3lept2}) remains valid but now $r_c$ of
Eq.~(\ref{eq:rc}) depends on ratios of Wilson coefficients {\it and}
the ratio of FCNC effective couplings $r_{e \mu} \equiv
|2d_{e\mu}d_{ee}^*/a_{e \mu}|$.
The analogue of Eq.~(\ref{eq:ratio3lept}) now reads
\begin{equation}
\label{eq:ratio3lept_bis}
\frac{B_{\mu \to 3e}}{B_{\tau \to 3\mu}} = 
\frac{ |a_{e\mu}|^2}{|a_{\mu \tau}|^2}  
\frac{I_{\mu \to 3 e}}{I_{\tau \to 3 \mu}} \ 
\displaystyle\frac{1 + 
\displaystyle\frac{r_c (r_{e \mu})}{I_{\mu \to 3 e}}}{1 + 
\displaystyle\frac{r_c (r_{\mu \tau})}{I_{\tau \to 3 \mu}}} \ .  
\end{equation}
If one could measure both $\Gamma_{\mu \to 3e}/\Gamma_{\mu \to e
\gamma}$ and $\Gamma_{\tau \to 3 \mu}/\Gamma_{\tau \to \mu \gamma}$
(which determine $r_c (r_{e \mu})$ and $r_c (r_{\mu \tau})$
respectively), then Eq.~(\ref{eq:ratio3lept_bis}) provides another way
to test the MLFV hypothesis, insensitive to specific model details.

\section{Disentangling operators: hadronic processes}
\label{sect:hadronic}

As data becomes available on different lepton flavor violating
processes it will be possible to begin disentangling the contributions
of different operators. The rates for $\mu \to e \gamma $ and $\mu\to
3e$ decays and for $\mu$-to-$e$ conversion in nuclei depend through
different linear combinations on the Wilson coefficients in the
effective lagrangian in Eq.~(\ref{eq:eff-lag}).   
The ratio $\Gamma_{\mu \to 3e} /\Gamma_{\mu \to e \gamma}$ is
sensitive to contact four-lepton operators already present in the
operator basis ($O_{4L}^{(i)}$) or induced by $O_{LL}^{(1,2)}$ via
$Z^0$ exchange (see Eq.~(\ref{eq:ratio3lept2})).  Similarly,
$\Gamma_{\mu \to e\text{~conversion}} /\Gamma_{\mu \to e \gamma}$ is
sensitive to the contact terms $O_{LL}^{(i)}$ involving quarks either
directly ($O_{LL}^{(3-5)}$) or via $Z^0$ exchange ($O_{LL}^{(1,2)}$).
In absence of contact operators the latter ratio is, 
\begin{equation}
\label{eq:mueconv}
\frac{\Gamma_{\mu \to e}^{A}}{\Gamma_{\mu\to e\gamma}} =  \pi \, D_A^2
\end{equation}
where we use the notation of Ref.~\cite{Kitano:2002mt} for the
dimensionless nucleus-dependent overlap integral $D_A$.  Typical
values for the overlap integral in the case of a light and heavy
nucleus are~\cite{Kitano:2002mt}: $D_{Al} \simeq 0.036$ and $D_{Au}
\simeq 0.19$.  A deviation from the result of Eq.~(\ref{eq:mueconv})
could be attributed to the contribution of $O_{LL}^{(i)}$.

After the cancellation of the MECO~\cite{MECO} experiment, which would
have had a single event sensitivity to branching fraction for
$\mu$-to-$e$ conversion of $2\times 10^{-17}$, it seems that
experimental searches of $\mu$-to-$e$ conversion are in the far
future.  This motivates us to explore the sensitivity of other
processes to $O_{LL}^{(3-5)}$.  In this section we study other decays
where hadrons are involved and $O_{LL}^{(3-5)}$ contribute at tree
level.
Finally, in Sec.~\ref{sec:no4L} we shall investigate whether
$O_{LL}^{(3-5)}$ can give significant contributions to $\mu \to3e$
through loop effects  in models where the 4L operators are not generated 
at the scale $\LLFV$ (and therefore their coefficients can be discarded).

\subsection{$\tau\to \pi^0\mu $}
Consider $\tau\to\mu X$, where $X=\pi^0$, $\rho^0$, etc. With a
non-vanishing left handed neutrino Majorana mass term, electroweak
interactions will produce a non-zero amplitude at one loop.  Roughly,
this amplitude corresponds to an operator
\beq
O_{SM}=\left[\frac{\alpha^2}{s_w^2M_W^2M_Z^2}\right]\,(\hat U m_\nu^2 \hat U^\dagger)_{\mu\tau} (\bar
\mu_L\gamma^\sigma \tau_L)(\bar q \gamma_\sigma(\gamma_5) q).
\eeq
This is much smaller than the contribution from the operators
generated at the LFV scale. For example, for minimal content the
factor in square brackets, of order $10^{-10} \,\text{GeV}^{-4}$, is
replaced by $1/v^4(\LLN/ \LLFV)^2\sim 10^{-9} \,\text{GeV}^{-4} (\LLN/
\LLFV)^2$, which for any reasonable value of the ratio $\LLN/ \LLFV$
is orders of magnitude bigger.

\begin{figure}[!t]
\centering
\epsfxsize=9cm
\epsffile{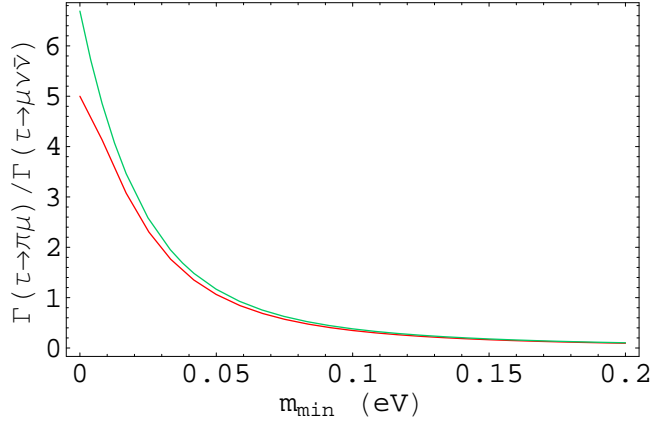}
\caption{\label{fig:taupimu-plot} Ratio $\Gamma(\tau\to\pi\mu)/ \Gamma(\tau\to\mu\nu\bar\nu)$ in
units of $10^{-27} (M_\nu v/ \LLFVsq)^2$, in the
extended field content scenario as a function of the lightest neutrino
mass (in eV). The red (lower) curve is for the normal hierarchy of neutrino
masses while the green (upper) curve is for the inverted hierarchy.}
\end{figure}

A straightforward calculation gives
\beq
\Gamma(\tau\to\mu\pi^0)=\frac{1}{256\pi}\frac{m_\tau^3
\Delta^2_{\mu\tau}f_\pi^2}{\LLFVfourth}
\big[
c^{(1)}_{LL} + c^{(2)}_{LL}
-c^{(4d)}_{LL}+c^{(4u)}_{LL}+2c^{(5)}_{LL}\big]^2,
\eeq
where $f_\pi=130\text{~MeV}$ is the $\pi$ decay constant. Numerically we
find, in the minimal model
\beq
\frac{\Gamma(\tau\to\mu\pi^0)}{\Gamma(\tau\to\mu\nu\bar\nu)}=
5.2\times10^{-52}\left(\frac{\LLN}{\LLFV}\right)^4
\big[
c^{(1)}_{LL} + c^{(2)}_{LL}
-c^{(4d)}_{LL}+c^{(4u)}_{LL}+2c^{(5)}_{LL}\big]^2.
\eeq
Using $\LLN/ \LLFV=10^9$ which corresponds to $\text{Br}(\mu\to
e\gamma)\sim10^{-13}$, and assuming the coefficients $c^{(i)}_{LL}$
are of order unity, this yields $\text{Br}(\tau\to\mu\pi^0)\sim
10^{-15}$.

For the extended model we obtain 
\beq
\frac{\Gamma(\tau\to\mu\pi^0)}{\Gamma(\tau\to\mu\nu\bar\nu)}=5.0\times10^{-27}
\left(\frac{vM_\nu}{\LLFVsq}\right)^2\big[
c^{(1)}_{LL} + c^{(2)}_{LL}
-c^{(4d)}_{LL}+
c^{(4u)}_{LL}+2c^{(5)}_{LL}\big]^2,
\eeq
in the normal hierarchy case, assuming $m_{\text{min}}=m_{\nu_1}\ll
m_{\nu_3}\approx \sqrt{\Delta m^2_{\text{atm}}}$. The result is
plotted as a function of the lightest neutrino mass, $m_{\text{min}}$,
in Fig.~\ref{fig:taupimu-plot} both for normal and inverted
hierarchies.  If $M_\nu =3\times10^5\LLFVsq/v$, which again
corresponds to $\text{Br}(\mu\to e\gamma)\sim10^{-13}$, we obtain
$\text{Br}(\tau\to\mu\pi^0)\sim 10^{-15}$. 

\subsection{$\pi^0\to\mu^+e^-$}
With almost no extra work we can compute
\beq
\Gamma(\pi^0\to\mu^+e^-)=\frac{f_\pi^2m_\mu^2(m_\pi^2-m_\mu^2)^2|
\Delta_{e\mu}|^2}{128\pi m_\pi^3 \LLFVfourth}
\big[
c^{(1)}_{LL} + c^{(2)}_{LL}
-c^{(4d)}_{LL}+c^{(4u)}_{LL}+2c^{(5)}_{LL}\big]^2\,.
\eeq
Numerically, using $1/\tau(\pi^0)=7.83$~eV, 
\beq
\text{Br}(\pi^0\to\mu^+e^-)= 1.4\times10^{-9}
\big[
c^{(1)}_{LL} + c^{(2)}_{LL}
-c^{(4d)}_{LL}+c^{(4u)}_{LL}+2c^{(5)}_{LL}\big]^2
\begin{cases}
(\LLN/ \LLFV)^4 |a_{e\mu}|^2 & \text{minimal}\\
(M_\nu v/ \LLFVsq)^2 |b_{e\mu}|^2 & \text{extended}
\end{cases}
\eeq
Assuming $c_{LL}^{(i)} \sim O(1)$ and $\LLN/ \LLFV=10^9$ one expects 
$\text{Br}(\pi^0\to\mu^+e^-) \sim 10^{-25}$.

\subsection{$J/ \psi\to\tau \mu$ and $\Upsilon\to \tau \mu$ }
The lepton flavor violating decays of heavier neutral mesons are 
interesting because decays into $\tau$'s become energetically 
allowed. By normalizing the decay rate for $V\to\ell\ell'$ (for $V=J/ \psi,
\Upsilon$) to the well measured $V\to ee$ we can make a prediction free
of hadronic uncertainties: 
\begin{equation}
\frac{\Gamma(V\to\ell_i\ell_j)}{\Gamma(V\to ee)} =
\frac1{e^2}\left(\frac{m_{V}}{2v}\right)^4
\big| \tilde{C}_V \big|^2
\,
\begin{cases}
(\LLN/ \LLFV)^4 |a_{ij}|^2 & \text{minimal}\\
(M_\nu v/ \LLFVsq)^2 |b_{ij}|^2 & \text{extended}
\end{cases} \ ,
\end{equation}
with 
\begin{equation}
\tilde{C}_V =  
\begin{cases}
\left(c^{(1)}_{LL} + c^{(2)}_{LL} \right)
\left( \frac{1}{2} - \frac{4}{3} \sin^2 \theta_w \right)      
+ c^{(3)}_{LL}+c^{(4u)}_{LL} - c^{(5)}_{LL} \qquad V=J/ \psi
\\
\left(c^{(1)}_{LL} + c^{(2)}_{LL} \right)
\left(-\frac{1}{2} + \frac{2}{3} \sin^2 \theta_w \right)
+c^{(3)}_{LL}+c^{(4d)}_{LL} + c^{(5)}_{LL} \qquad V=\Upsilon
\end{cases} \ .
\end{equation}
In particular, for $\Upsilon$ decays into $\tau\mu$ this gives
\begin{align}
\text{Br}(\Upsilon \to\tau^+\mu^-)=&
\left[
\left(c^{(1)}_{LL} + c^{(2)}_{LL} \right)
\left(-\frac{1}{2} + \frac{2}{3} \sin^2 \theta_w \right) +
c^{(3)}_{LL}+c^{(4d)}_{LL}+c^{(5)}_{LL}\right]^2  
\, \times \\
\times & 
\begin{cases}
3.2\times10^{-55}(\LLN/ \LLFV)^4 & \text{minimal}\\
3.2\times10^{-30}(M_\nu v/ \LLFVsq)^2 & \text{extended, normal h,
$m_{\nu\text{min}}=0$,}\\
4.3\times10^{-30}(M_\nu v/ \LLFVsq)^2 & \text{extended, inverted h,
$m_{\nu\text{min}}=0$.}
\end{cases}
\end{align}
Assuming again $c_{LL}^{(i)} \sim O(1)$ and $\LLN/ \LLFV=10^9$ one
expects $\text{Br}(\Upsilon \to \tau^+ \mu^-) \sim 10^{-20}$. 

All the hadronic modes we have considered here, for acceptable values
of the ratios $\LLN/\LLFV$ and $v M_\nu/\LLFV^2$ and of the Wilson
coefficients, are predicted to be well below foreseeable experimental
sensitivities~\cite{danko}. On one hand this result can be considered
as a generic prediction of minimal lepton flavor violation and can be
used in the future to test this framework (e.g. observation of
$\Upsilon \to\tau^+\mu^-$ at B factories would strongly disfavor it).
On the other hand we see that within minimal lepton flavor violation
the most sensitive probes of $O_{LL}^{(3-5)}$ are: (i) $\mu \to e$
conversion in nuclei and (ii) purely leptonic processes such as $\mu
\to 3 e$ (to which $O_{LL}^{(3-5)}$ contribute via loops).

\section{$\mu\to eee$ in the absence of 4L operators}
\label{sec:no4L}

The operators $O^{(3-5)}_{LL}$ can give rise to $\mu\to eee$ via one
loop graphs. These contributions can be significant in models where
the 4L operators are not generated at the scale $\LLFV$, or are
generated with tiny Wilson coefficients.  We will analyze now this
dynamical scenario, which amounts to the assumption $c_{4L}^{(i)}
(\LLFV) \ll c_{LL}^{(j)} (\LLFV)$.  In order to address the
sensitivity of $\mu\to eee$ to $c^{(3-5)}_{LL}$, we organize our
analysis as follows: first we use the renormalization group to evolve
the effective lagrangian from $\mu \sim \LLFV$ down to $\mu \sim 1$
GeV.  Then we take $\mu \to 3e$ matrix elements of the relevant
operators, using perturbative QCD to deal with the heavy
quarks and chiral perturbation theory to deal with the light quark
loops.

\subsection{RG analysis} 
\label{sec-RG}

We use the renormalization group equations (RGEs) to determine the low
energy effective lagrangian. This is done is three steps: first, the
RGE in the unbroken phase of the $SU(2)\times U(1)$ theory is used to
compute the coefficients in the effective lagrangian in
Eq.~(\ref{eq:eff-lag}) down to a scale $\mu\sim M_Z$. In the second
step the coefficients are matched to those of an effective lagrangian
for the theory in the broken symmetry phase of $SU(2)\times U(1)$. And
third, the coefficients of this effective lagrangian are computed at
$\mu\sim1$~GeV using the RGE for the theory with only $U(1)$ gauge
group. The calculation is rather lengthy and technical, so the details
are presented in Appendix~\ref{app-RG}.

In the first step of the calculation we compute for $M_Z < \mu <
\LLFV$ the mixing of coefficients $c^{(i)}_{LL}$ into the coefficients
of the four-lepton contact interactions, $c^{(j)}_{4L}$, assuming
$c_{4L}(\LLFV)\ll c_{LL}(\LLFV)$. In the second step, an effective
theory with $U(1)$ gauge field is constructed with effective
lagrangian given by matching at $\mu\sim M_Z$ to the theory with
$SU(2)\times U(1)$ symmetry. The relevant operators in the new effective
theory are
\begin{equation}
\label{eq:ops-matched}
\begin{aligned} 
\hat O^{(3)}_{LL} &= \bar e_L\gamma^\mu \XLFV e_L\; (\bar u_L\gamma_\mu u_L+\bar d_L\gamma_\mu d_L) \\
\hat O^{(4d)}_{LL} &= \bar e_L\gamma^\mu \XLFV e_L\; \bar d_R\gamma_\mu d_R \\
\hat O^{(4u)}_{LL} &= \bar e_L\gamma^\mu \XLFV e_L\; \bar u_R\gamma_\mu u_R \\
\hat O^{(5)}_{LL} &= -\bar e_L\gamma^\mu \XLFV e_L\;(\bar u_L\gamma_\mu u_L-\bar d_L\gamma_\mu d_L)\\
\hat O^{(1)}_{RL} & = ev \bar e_R\sigma^{\mu\nu} \lambda_e^{\phantom{\dagger}}\XLFV e_L\, F_{\mu\nu}\\
\hat O^{(1)}_{4L} & =\bar e_L\gamma^\mu \XLFV e_L\; \bar e_L\gamma_\mu e_L \\
\hat O^{(3)}_{4L} & =\bar e_L\gamma^\mu \XLFV e_L\; \bar e_R\gamma_\mu e_R \\
\end{aligned} 
\end{equation}
where sums over quark generations are understood (the top quark has been
integrated out), and the effective lagrangian is  
\begin{equation}
\label{eq:eff-lag-low}
\CL=\frac{1}{\LLFVsq}
\sum_{i}\Big(\hat c_{LL}^{(i)} \hat O_{LL}^{(i)} +\hat c_{4L}^{(i)} \hat
O_{4L}^{(i)}\Big)+   
\frac{1}{\LLFVsq} \left( \sum_{j} \hat c_{RL}^{(j)} \hat O_{RL}^{(j)}  +\hc  \right)
\end{equation}
The coefficients $\hat{c}_{LL,RL,4L}$ , obtained by matching, are given in
Eq.~(\ref{eq:matching1}) of Appendix~\ref{app-RG}. 

Finally, in the third step we solve the RGE satisfied by 
$\hat{c}^{(1)}_{4L}$  and $\hat{c}^{(3)}_{4L}$. 
Ignoring the running of $c_{LL/RL}$ (that is working to order $g^2$),
we finally obtain the following low scale ($\mu < M_Z$) Wilson
coefficients:
\begin{equation}
\begin{aligned}
\hat c^{(1)}_{4L}(\mu)&=
-\frac{\alpha}{\pi}\left\{\frac13\ln\left(\frac{\mu}{M_Z}\right)\left[
c^{(3)}_{LL}+4c^{(4u)}_{LL}-3 c^{(4d)}_{LL}-7
c^{(5)}_{LL}\right]\right.\\
&\qquad\left. 
+\frac{2}{3}\ln\left(\frac{\sqrt{2} m_t}{M_Z}\right)\left[
c^{(3)}_{LL}+c^{(4u)}_{LL} -c^{(5)}_{LL} \right]
\right.  \\
& \qquad\left. +\ln\left(\frac{M_Z}{\LLFV}\right)\left[
\frac{1}{2\cos^2\theta_w}\left(c^{(3)}_{LL}+2c^{(4u)}_{LL}-
c^{(4d)}_{LL}\right)
-\frac{3}{2\sin^2\theta_w}c^{(5)}_{LL}\right]\right\}\\ 
& \qquad\left.   
- \left(\frac{1}{2} - \sin^2\theta_w \right) 
\left( c^{(1)}_{LL} + c^{(1)}_{LL} \right)~,
\right. \\
\hat c^{(3)}_{4L}(\mu)&=
 -\frac{\alpha}{\pi}\left\{\frac13\ln\left(\frac{\mu}{M_Z}\right)\left[
c^{(3)}_{LL}+4c^{(4u)}_{LL}-3 c^{(4d)}_{LL}-7
c^{(5)}_{LL}\right]\right.\\
&\qquad\left. 
+ \frac{2}{3}\ln\left(\frac{\sqrt{2} m_t}{M_Z}\right)\left[
c^{(3)}_{LL}+c^{(4u)}_{LL} -c^{(5)}_{LL}\right]
\right.  \\
&\qquad\left.
+\ln\left(\frac{M_Z}{\LLFV}\right)\left[
\frac{1}{\cos^2\theta_w}\left(c^{(3)}_{LL}+2c^{(4u)}_{LL}-
c^{(4d)}_{LL}\right)\right]\right\} 
+ \sin^2\theta_w  \left( c^{(1)}_{LL} + c^{(1)}_{LL} \right)~.
\\
\end{aligned}
\label{eq:wc4L}
\end{equation}
%

\subsection{Matrix elements}

At the order in $g^2$ we are working, the matrix elements of
$\hat{O}_{4L}^{(1,3)}$ have to be taken at tree level, while those of
$\hat{O}^{(i)}_{LL}$ have to be evaluated at one loop.  Consider for
definiteness the decay $ \mu^- (p_\mu) \to e^- (p_1) \ e^- (p_2) \ e^+
(p_{\bar e})$.  Let us define:
$$ {\cal M}_{\mu \nu} (p_1, p_2) = 
\Delta_{e \mu} \bar{u} (p_1) \gamma_\mu P_L u(p_\mu) \cdot 
\bar{u}(p_2) \gamma_\nu v (p_{\bar e}) \qquad \quad 
{\cal M} (p_1, p_2) =  g^{\mu \nu}  {\cal M}_{\mu \nu} (p_1, p_2)~, 
$$
and the kinematic variables 
$x=m_{13}^2/m_\mu^2=(p_{e_1}+p_{\bar e})^2/m_\mu^2$ and
$y=m_{23}^2/m_\mu^2=(p_{e_2}+p_{\bar e})^2/m_\mu^2$.\\

\begin{figure}[t]
\centering
\epsfxsize=6cm
\epsffile{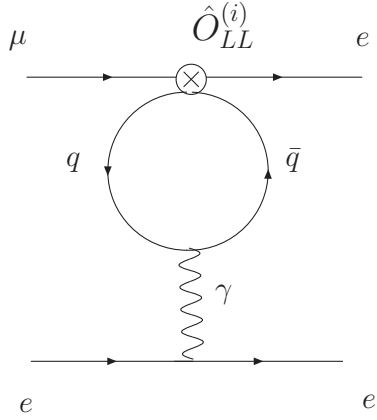}
\caption{\label{fig:hq}
One-loop contribution to $\langle 3 e | \hat{O}_{LL}^{(i)}| \mu \rangle$. 
The heavy quark contribution is treated in perturbation theory, 
while the light quark contribution is evaluated using 
chiral perturbation theory. 
}
\end{figure}

\noindent
{\bf \underline{Matrix 
elements of $\hat{O}^{(i)}_{LL}$: heavy quark contributions}} \\
Considering the graphs of Fig.~\ref{fig:hq} with heavy quark internal
loops, we find
\begin{equation}
\begin{aligned} 
\langle 3e | \sum_{i} \hat{c}^{(i)}_{LL} \hat{O}^{(i)}_{LL} | \mu \rangle  
= &  -\frac{\alpha}{\pi} \Big[
(2 F_c (y)  - F_b (y) ) \hat{c}^{(3)}_{LL} +
 2 F_c (y)   \hat{c}^{(4u)}_{LL} 
\\
& - \ F_b (y)  \hat{c}^{(4d)}_{LL} -
(2 F_c (y)  + F_b (y) )  \hat{c}
^{(5)}_{LL} 
\Big] \times {\cal M}(p_1,p_2) 
\\
&  - \ \Big\{ p_1  \leftrightarrow p_2 \ ; \ 
y  \leftrightarrow x \Big\}~ , 
\end{aligned}
\end{equation}
where  $F_q(x)\equiv F(m_q,(m_q/m_\mu)^2/x)$, with
\beq
\label{eq:Ffunction}
F(m,z)=
\begin{cases}
\frac16\ln\frac{2m^2}{\mu^2}-\frac5{18}-\frac{2z}3+\frac13(1+2z)\sqrt{4z-1}\,\arctan\left(\frac1{\sqrt{4z-1}}\right)
&\text{for $4z>1$}\\
\frac16\ln\frac{2m^2}{\mu^2}-\frac5{18}-\frac{2z}3+
+\frac16(1+2z)\sqrt{1-4z}\,\left[\ln\left(\frac{1+\sqrt{1-4z}}{1-\sqrt{1-4z}}\right)+i\pi\right]
&\text{for $4z<1$}
\end{cases}
\eeq
In muon decays $m_{13}^2, m_{12}^2 \sim q^2 < m_\mu^2$,  
and by using 
$$ F (m,m^2/q^2) = \frac{1}{6} \log \frac{ 2 m^2}{\mu^2} - 
\frac{q^2}{30 m^2} + \dots $$ 
the heavy quark contribution can be safely approximated to a constant
(local) term.   
Using the $ q^2/m_q^2 \to 0 $ limit and the notation $F_q  \equiv  1/3 \log
\left[ (\sqrt{2} m_q)/\mu \right]$ we then obtain: 
\begin{equation}
\begin{aligned}
 \langle 3e | \sum_{i} \hat{c}^{(i)}_{LL} \hat{O}^{(i)}_{LL} | \mu \rangle  
= & 
-\frac{\alpha}{\pi} \left[ 
(2 F_c  - F_b) \hat{c}^{(3)}_{LL} +
 2 F_c  \hat{c}^{(4u)}_{LL} 
- F_b  \hat{c}^{(4d)}_{LL} -
(2 F_c + F_b)  \hat{c}^{(5)}_{LL} 
\right] \\
&  \times \left[ 
\langle \hat{O}_{4L}^{(1)} \rangle + \langle \hat{O}_{4L}^{(3)} \rangle
\right] 
\end{aligned}
\end{equation}
It is easy to check that the scale dependence of the above matrix
elements cancels the one induced by $b,c$ loops in the Wilson
coefficients $\hat{c}_{4L}^{(1)},\hat{c}_{4L}^{(3)}$  
given in Eq.~(\ref{eq:wc4L}). \\

\begin{figure}[t]
\centering
\epsfxsize=10cm
\epsffile{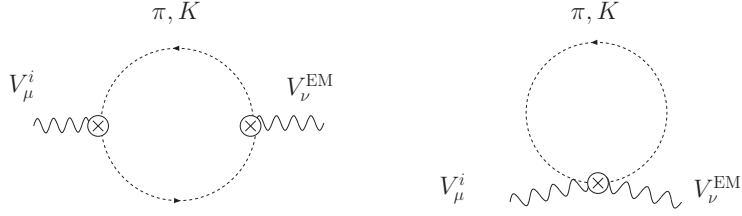}
\caption{\label{fig:chiralloops}
One-loop contributions to the VV correlator 
defined in Eq.~(\ref{eq:VVchpt}) in $SU(3)\times SU(3)$ chiral perturbation 
theory. The vertices come from the lowest order ($O(p^2)$) chiral lagrangian  
coupled to external sources.  
}
\end{figure}

\noindent
{\bf \underline{Matrix elements of $\hat{O}^{(i)}_{LL}$: 
light quark contributions}}\\ 
The operators $\hat{O}_{LL}^{(i)}$ have the structure
\begin{equation}
\hat{O}_{LL}^{(i)} = 
\bar e_L\gamma^\mu \XLFV e_L \ \frac{1}{2} \Big[ 
V_{\mu}^{(i)}  \pm A_{\mu}^{(i)} 
\Big]~,
\end{equation}
where the flavor structure of the currents 
$V_\mu^{(i)}$, $A_\mu^{(i)}$ can be read off Eq.~(\ref{eq:ops-matched}). 
Only the vector current contributes to the matrix elements in
question and we find: 
\begin{equation}
\begin{aligned}
 \langle 3e |  \hat{O}^{(i)}_{LL} | \mu \rangle  
= & - i \frac{e^2}{2}  \, {\cal M}^{\mu \nu} (p_1,p_2) \, 
\frac{1}{(p_2 + p_{\bar{e}} )^2} \times 
\\
\times & 
\int d^4x  \, e^{i x (p_2 + p_{\bar{e}})}  
\langle 0 | T \Big( V_\mu^i (0)\, V_{\nu}^{\rm EM}(x) \Big)  |0 \rangle 
\ - \ \big\{p_1 \leftrightarrow p_2 \big\} 
\end{aligned}
\end{equation}
Defining then 
\begin{equation}
\label{eq:VVchpt}
\int d^4x  \, e^{i x q }  
\langle 0 | T \Big( V_\mu^i (0)\, V_{\nu}^{\rm EM}(x) \Big)  |0 \rangle =
\frac{i}{(4 \pi)^2} \left( g_{\mu \nu} q^2 - q_\mu q_\nu \right) \cdot 
\Pi^{i,{\rm EM}} (q^2) \ , 
\end{equation}
one arrives at: 
\begin{equation}
\langle 3e |  \hat{O}^{(i)}_{LL} | \mu \rangle =
\frac{e^2}{2 (4 \pi)^2}  \, {\cal M} (p_1,p_2) \, 
\Pi^{i,{\rm EM}} \left( (p_2 + p_{\bar{e}})^2 \right)  
\ -  \ \big\{p_1 \leftrightarrow p_2 \big\} 
\label{eq:me1}
\end{equation}
In order to evaluate the VV correlator at low momentum transfer we use
$SU(3) \times SU(3)$ chiral perturbation theory~\cite{gl85} to 
order $p^4$.  
Evaluating the one-loop diagrams depicted in Fig.~\ref{fig:chiralloops} 
(with vertices from the $O(p^2)$ chiral lagrangian) and adding the counterterm
contributions from the local $O(p^4)$ effective lagrangian~\cite{gl85}, 
we find:
\begin{eqnarray}
\Pi^{4u,{\rm EM}} \left( q^2 \right) &=& 
\left[ \sum_{i=\pi,K}   
f\left(m_i^2/{q^2}, m_i^2/\mu_\chi^2 \right)  - (4 \pi)^2 
\frac{8}{3} \left( 2 H_1^r (\mu_\chi,\mu) + L_{10}^r (\mu_\chi) \right)
\right]
\nonumber \\
f(z,m^2/\mu_\chi^2)&=&
-\frac13\ln\frac{m^2}{\mu_\chi^2}+\frac89-\frac83z
+\frac23(4z-1)^{3/2}\cot^{-1}(\sqrt{4z-1})\,.
\label{eq:orrchpt}
\end{eqnarray}
The effective couplings $L_{10}$ and $H_1$ are defined in
Ref.~\cite{gl85}, and they cancel the loop-induced dependence on the
chiral renormalization scale $\mu_\chi$.  The constant $H_1$, however,
depends on the renormalization scheme adopted in the short distance
theory (at the quark level), which is in this case the $\overline{\rm
MS}$ scheme.  This dependence is such as to cancel the $\mu$
dependence of the amplitude induced by the Wilson coefficients (see
appendix~\ref{app:chpt} for details).  Similarly we find $\Pi^{3,{\rm
EM}} \left( q^2 \right) = 0$, $\Pi^{4d,{\rm EM}} \left( q^2 \right) =
- \Pi^{4u,{\rm EM}} \left( q^2 \right)$, and $\Pi^{5,{\rm EM}} \left(
q^2 \right) = 2 \Pi^{4u,{\rm EM}} \left( q^2 \right)$.  The final
result for the light-quark loop matrix elements is
\begin{equation}
\begin{aligned}
 \langle 3e | \sum_{i} \hat{c}^{(i)}_{LL} \hat{O}^{(i)}_{LL} | \mu \rangle  
= & \
\frac{\alpha}{4 \pi} \, \frac{1}{2} 
\left[ \hat{c}^{(4u)}_{LL} - \hat{c}^{(4d)}_{LL} +
 2 \hat{c}^{(5)}_{LL} \right] \times
\\
\times & \ {\cal M} (p_1,p_2) \ 
\Pi^{4u,{\rm EM}} \left( (p_2 + p_{\bar{e}})^2 \right)  
\ -  \ \big\{p_1 \leftrightarrow p_2 \big\}~.
\end{aligned}
\end{equation}
Over the physical region for $\mu\to3e$ decay the function
$\Pi^{4u,{\rm EM}} \left( q^2 \right)$ varies 
by less than 1\%~\footnote{ 
For the low energy constants we use 
$L_{10}^r(\mu_\chi=m_\rho) = - 5.5 \times 10^{-3}$ (experiment, see e.g.
\cite{Pich:2002xy}) and 
$|H_{1}^r(\mu_\chi=m_\rho, \mu = 1 \, {\rm GeV})| \leq  1/(64 \pi^2)$ 
(naive dimensional analysis~\cite{Manohar:1983md} bound). 
}.  
Approximating it to a constant we find:
\begin{equation}
\langle 3e | \sum_{i} \hat{c}^{(i)}_{LL} \hat{O}^{(i)}_{LL} | \mu \rangle  
= 
\frac{\alpha}{4 \pi} \, \frac{1}{2} 
\left[ \hat{c}^{(4u)}_{LL} - \hat{c}^{(4d)}_{LL} +
 2 \hat{c}^{(5)}_{LL} \right] \ 
\Pi^{4u,{\rm EM}} \left( m_\mu^2 \right) \times 
\left[ 
\langle \hat{O}_{4L}^{(1)} \rangle + \langle \hat{O}_{4L}^{(3)} \rangle
\right]~.
\end{equation}

In summary, the matrix element calculation results in:
\begin{equation}
\label{eq:mesummary}
\langle 3e | \sum_{i} \hat{c}^{(i)}_{LL} \hat{O}^{(i)}_{LL} | \mu \rangle  
= 
\left( \kappa_{\rm heavy-q} + \kappa_{\rm light-q} \right) 
\times \left( 
\langle \hat{O}_{4L}^{(1)} \rangle + \langle \hat{O}_{4L}^{(3)} \rangle
\right)~, 
\end{equation}
with  
\begin{equation}
\begin{aligned}
\kappa_{\rm heavy-q} = & 
-\frac{\alpha}{\pi} \left[ 
(2 F_c  - F_b) \hat{c}^{(3)}_{LL} +
2 F_c  \hat{c}^{(4u)}_{LL} 
- F_b  \hat{c}^{(4d)}_{LL} -
(2 F_c + F_b)  \hat{c}^{(5)}_{LL} 
\right] \\
\kappa_{\rm light-q} = &
\frac{\alpha}{8 \pi} \,
\left[ \hat{c}^{(4u)}_{LL} - \hat{c}^{(4d)}_{LL} +
 2 \hat{c}^{(5)}_{LL} \right]  
\Pi^{4u,{\rm EM}} \left( m_\mu^2 \right) \\
\end{aligned}
\end{equation}

\subsection{Rate}

The result in Eq.~(\ref{eq:mesummary}) implies that in the scenario
considered in this section the $\mu \to e \bar{e} e$ rate is given in
closed form by Eq.~(\ref{eq:rate1}), replacing  the coefficients $a_{\pm}$
of Eq.~(\ref{eq:ai}) by:
\begin{equation}
\begin{aligned}
a_+ =& \hat{c}^{(3)}_{4L} 
+ \kappa_{\rm heavy-q} 
+ \kappa_{\rm light-q} \\
a_- =& \hat{c}^{(1)}_{4L} 
+ \kappa_{\rm heavy-q} 
+ \kappa_{\rm light-q} \\
\end{aligned}
\ , 
\end{equation}
with $\hat{c}^{(1,3)}_{4L}$ given by Eq.~(\ref{eq:wc4L}).
Taking into account the $\mu$ dependence of $H_1^r(\mu_\chi,\mu)$ it is 
straightforward to verify that $a_{\pm}$ do not depend on the 
renormalization scale $\mu$.  

While the formulas given above are quite general, 
in order to illustrate at which level $\mu \to e \bar{e} e$ decays
probe $O_{LL}^{(3-5)}$, let us assume that only one
operator dominates. Moreover, let us focus on the operator
contributing with the largest numerical coefficient, which turns out
to be $O_{LL}^{(5)}$. We find then: 
\begin{equation}
\label{eq:loop-rate}
\Gamma_{\mu \to 3e}/ \Gamma_{\mu \to e \nu \bar{\nu}} =  6 \cdot 10^{-3}
\times \big| c_{LL}^{(5)} (\LLFV) \big|^2  \times 
\begin{cases} \left(\frac{\LLN}{\LLFV}\right)^4 |a_{e \mu}|^2 
& \text{minimal field content}\\
\left(\frac{v M_\nu}{\LLFV^2}\right)^2 |b_{e \mu}|^2 
& \text{extended field content}\end{cases} ~ . 
\end{equation}
A comparison of Eqs.~(\ref{eq:loop-rate}) and (\ref{eq:rate3}) shows
that the rates of $\ell\to\ell'\ell''\bar\ell'''$ decays in the
scenario $c_{4L}^{(i)} (\LLFV) \ll c_{LL}^{(j)} (\LLFV)$ are typically
suppressed relative to their tree level counterparts by
$\sim10^{-3}$. This factor is larger than the naive estimate $(\alpha
/\pi)^2$ because of the presence of sizeable logarithms.  Similar
logarithmic enhancements were first pointed out within a non-MLFV
effective theory approach to lepton flavor violation in
Ref.~\cite{Raidal:1997hq} (in that reference the main focus
was on the one-loop contributions to $\mu$-to-$e$ conversion in
nuclei). Finally, let us notice that apart from the overall
suppression, in this dynamical scenario we still find the pattern
$\Gamma(\tau \to \mu) \gg \Gamma(\mu \to e) \sim \Gamma(\tau \to e)$,
which is dictated by the MLFV hypothesis.

\section{Conclusions}
\label{sect:conclusions}
The MLFV hypothesis provides a framework for discussing and analyzing
the predictions of a large class of models where Lepton Flavor
Violation arises solely from lepton mass matrices. The framework is
attractive because it is both very general and fairly predictive. We
have extended the results of Ref.~\cite{Cirigliano:2005ck}, which
introduced the MLFV hypothesis and analyzed $\mu$-to-$e$ conversion in
Nuclei and radiative decays, e.g., $\mu\to e \gamma$, to decays
involving hadrons, e.g., $\tau\to\mu\pi$, and to decays involving only charged
leptons, e.g., $\mu\to3e$.  To this end we extended the operator basis for
the effective lagrangian introduced in \cite{Cirigliano:2005ck}. In
particular, we found five new purely leptonic operators,  given in
Eq.~(\ref{eq:4Lops}), which were omitted in \cite{Cirigliano:2005ck}
because they do not contribute directly to the processes considered
there.

In the event that several LFV processes are observed and their rates
measured one could begin to test the MLFV hypothesis since it predicts
definite patterns of relative magnitudes of rates. We have therefore
computed the rates for the hadronic processes $\tau\to\ell\pi$
($\ell=\mu, e$), $\pi^0\to\mu^+e^-$ and $V\to\tau\mu$ ($V=J/ \psi,
\Upsilon $), and for purely charged lepton decays $\mu\to3e$,
$\tau\to3\ell $ ($\ell=e,\mu $), $\tau\to \bar\ell \ell \ell'$ and
$\tau\to \ell \ell \bar \ell'$ ($(\ell,\ell')=(e,\mu), (\mu,e)$). One
definite prediction of the MLFV hypothesis is that the rates for
decays involving hadrons are exceedingly small. For example, if the
scales of LFV and LNV are such that $\text{Br}(\mu\to
e\gamma)\sim10^{-13}$, then $\text{Br}(\tau\to\mu\pi^0)\sim10^{-15}$,
$\text{Br}(\pi^0\to\mu^+e^-)\sim10^{-25}$ and
$\text{Br}(\Upsilon\to\tau^+\mu^-)\sim10^{-20}$, well below the
sensitivity of foreseeable experiments. 

We analyzed the decays involving only charged leptons first at tree
level and then through loop processes, in order to investigate the
contribution of the operators involving quarks (under the assumptions
that the coefficients of the 4-lepton operators of
Eq.~(\ref{eq:4Lops}) are negligibly small).  Some general conclusions
apply to both cases: (i) LFV $3\ell$ decays display the same pattern
$B_{\tau\to\mu}\gg B_{\mu\to e}\sim B_{\tau\to e}$ as the radiative
decays, and hence offer an alternative way to test the MLFV
hypothesis; (ii) ratios of transitions between the same two families,
as, for example $\Gamma(\mu\to3e)/\Gamma(\mu\to e\gamma)$, are
determined by ratios of various Wilson coefficients and therefore the
combined measurement would be valuable in disentangling the
contributions of different operators.

More specifically, the tree level rates for decays involving only
charged leptons can be very significant. If the coefficients in the
effective lagrangian are of order 1 and the ratio of scales of LNV and
LFV is as large as allowed by the condition that the Yukawa couplings
are perturbative, then in the case of minimal field content
$\text{Br}(\mu\to3e)$ could easily exceed the present 90\% C.L. limit
of $1.0\times10^{-12}$. In the extended field content case, under the
same assumptions, this branching fraction is comparable to the current
experimental limit, and the fractions for LFV $\tau$ decays are three
orders of magnitude smaller than the current limits of
$1-3\times10^{-7}$\cite{Aubert:2003pc,Yusa:2004gm}. Note that while
the rates for LFV $\tau$ decays are larger than for $\mu\to3e$, the
higher experimental sensitivity of the latter is on the verge of
placing stringent bounds on the theory and therefore it would be
highly desirable to pursue higher sensitivity in this mode.

If the new physics giving rise to the effective lagrangian at the LFV
scale is tied to the quark sector in such a way that only operators
involving quarks and leptons together were produced by this dynamics,
then the decays to $3\ell$ would not proceed at tree level. To
investigate this scenario we computed the rates for $3\ell$ decays
through loops, assuming the coefficients of the 4-lepton operators of
Eq.~(\ref{eq:4Lops}) are negligibly small. Alternatively, since the
rates for processes involving hadrons are exceedingly small, a more
sensitive probe of them may arise through their loop effect on $3\ell$ decays.
Not surprisingly we found the rates are suppressed relative to their
tree level counterparts by $\sim10^{-3}$. The calculation is interesting
in its own right. It involves the computation of a low energy effective 
lagrangian obtained by integrating out the heavy quarks and the $W^\pm$
and $Z^0$ vector bosons, and also requires the use of the chiral
lagrangian to properly describe the low energy physics involving light
quarks. As a side result a new sum rule for the Gasser-Leutwyler
counter-term $H^r_1$ was derived in  Appendix~\ref{app:chpt}. 

A discrimination of various dynamical scenarios within MLFV will be
possible only when more than one LFV decay mode is observed.  This
remains true even when analyzing data beyond the minimal flavor
violation hypothesis~\cite{review}.  We therefore emphasize that it is
highly desirable to complement existing experimental
efforts~\cite{MEG,Aubert:2003pc,Yusa:2004gm} and pursue experimental
searches of {\it all} the LFV $\mu$ and $\tau$ decay modes.

\bigskip

\paragraph{Acknowledgments}

We thank Gino Isidori and Mark Wise for useful discussions and
collaboration in an early stage of this work. We are grateful to 
Valentina Porretti for help in producing scatter-plots. 
V.C.~was supported by Caltech through the Sherman Fairchild fund 
and acknowledges the hospitality of the INFN-LNF during the Fall 2005. 
This work was supported by the U.S.~Department of Energy  
under grants DE-FG03-97ER40546 and DE-FG03-92ER40701.

\appendix

\section{Differential distributions and phase space for 
$\ell\to\ell'\ell''\bar\ell'''$}
\label{sect:app1}

The differential rate for $\mu\to ee\bar e$ in the variables
$x=m_{13}^2/m_\mu^2=(p_{e_1}+p_{\bar e})^2/m_\mu^2$ and
  $y=m_{23}^2/m_\mu^2=(p_{e_2}+p_{\bar e})^2/m_\mu^2$  is:
\begin{multline}
\frac{d^2\Gamma}{dxdy}=\frac{m_\mu^5}{2^9\pi^3\LLFVfourth}\frac1{xy}\left[
2\left(2 x^3-2 x^2+x+2 y^3-2 y^2+y\right) |a_0|^2\right.\\
+|a_-|^2\left(-8 y x^3-16
   y^2 x^2+8 y x^2-8 y^3 x+8 y^2 x\right)
 +|a_+|^2 \left(-2 y x^3+2 y x^2-2 y^3 x+2 y^2
   x\right) \\
+a_0 \left(\left(8 y x^2+8 y^2 x-8 y
   x\right) a_-^{*}+\left(-2 y x^2-2 y^2 x\right)
   a_+^{*}\right)\\
\left.+\left(a_+ \left(-2 y x^2-2 y^2
   x\right)+a_- \left(8 y x^2+8 y^2 x-8 y x\right)\right)
   a_0^{*}\right]
\end{multline}
The phase space integral $I$ appearing in Eq.~\ref{eq:rate1} reads:
\begin{equation}
\label{eq:Iintegral}
I=\int_{4\hat m^2}^{(1-\hat
m)^2}\!\!\!\!\!\!\!\!\!dx\int_{y_{\text{min}}}^{y_{\text{max}}}\!\!\!\!\!
dy \left\{\frac{1}{xy}
\left[y(1-x)+x(1-y)-2(1-x-y)(x^2+y^2)\right]+2(1-x-y)\right\}
\end{equation}
in terms of  $x=(p_{e_1}+p_{\bar
e})^2/m_\mu^2$, $y=(p_{e_2}+p_{\bar e})^2/m_\mu^2$,
and $\hat m\equiv m_e/m_\mu$. The integration limits are given by:
\begin{equation}
y_{\substack{\text{min}\\ 
\text{max}}}=\frac12\left[1-x+3\hat m^2\mp\sqrt{\frac{(x-4\hat
      m^2)[(1-x-\hat m^2)^2-4\hat m^2 x]}{x}}\right]~.
\end{equation}
Numerically, $I=9.886$ for $\mu\to ee\bar e$ $I=17.4$ for $\tau \to
ee\bar e$ and $I=3.26$ for $\tau\to\mu\mu\bar\mu$.  An exact analytic
expression for $I$ cannot be readily obtained, but by expanding in
powers of $\hat m$ the
integrand in (\ref{eq:Iintegral}) after performing the $y$-integral,
keeping the full dependence on $\hat m$ in the limits of
$x$-integration gives the result in Eq.~(\ref{eq:Ips}), 
which numerically gives 10.4 for $\mu\to ee\bar e$.

The differential rate for $\tau^-\to e^-\mu^-\mu^+$ in the variables
$x=m_{12}^2/m_\tau^2=(p_{e}+p_{\mu^-})^2/m_\tau^2$ and
  $y=m_{23}^2/m_\tau^2=(p_{\mu^-}+p_{\mu^+})^2/m_\tau^2$  is
\begin{multline}
\frac{d^2\Gamma}{dxdy}=\frac{m_\tau^5}{2^8\pi^3\LLFVfourth}\frac1{y}\left[
2\left(2 x^2+2 y x-2 x-y+1\right) |a_0|^2+ \left(2 x y-2 x^2
y\right)|a_-|^2 \right.\\
  +\left(-2 y^3-4 x y^2+2 y^2-2 x^2 y+2 x y\right) |a_+|^2
  +a_0 \left(\left(2 y^2+2 x y-2 y\right) a_+^{*}-2 x y  a_-^{*}\right)\\
\left.+\left(a_+ \left(2 y^2+2 x y-2
   y\right)-2 a_- x y\right) a_0^{*}
\right]
\end{multline}
The integral over phase space $\tilde I$, appearing in Eq.~(\ref{eq:rate2}), 
is given by ($\hat m=m_\mu/m_\tau$)
\begin{equation}
\tilde I =\int_{\hat m^2}^{(1-\hat m)^2} dx \int_{\hat y_{\text{min}}}^{\hat
 y_{\text{max}}} dy~ \frac{1-y-2x(1-x-y)}{y} ~ .
\end{equation}
The numerical value is 1.50 for $\tau^-\to e^-\mu^-\mu^+$ and  
8.49 for the corresponding integral in the case of $\tau^-\to
\mu^-e^-e^+$. Note that the limits of integration are different in these
cases than in $\mu\to ee\bar e$.

\section{Low Energy Effective Lagrangian}
\label{app-RG}

In this appendix we give the details of the calculation described in
Sec.~\ref{sec-RG}, that is, the use of the renormalization group
equations to determine the low energy effective lagrangian. As
explained there, this is done is three steps: first, the RGE in the
unbroken phase of the $SU(2)\times U(1)$ theory is used to compute the
coefficients in the effective lagrangian in Eq.~(\ref{eq:eff-lag})
down to a scale $\mu\sim M_Z$. In the second step the coefficients are
matched to those of an effective lagrangian for the theory in the
broken symmetry phase of $SU(2)\times U(1)$. And third, the
coefficients of this effective lagrangian are computed at
$\mu\sim1$~GeV using the RGE for the theory with only $U(1)$ gauge
group.

In the region $\mu\gsim M_Z$ (neglecting $SU(2)\times U(1)$ symmetry 
breaking) a one loop computation gives
\begin{equation}
\begin{aligned} 
\mu\frac{dc^{(1)}_{4L}}{d\mu}&=-\frac1{18} N_cN_g\frac{g_1^2}{4\pi^2}[c^{(3)}_{LL}+2c^{(4u)}_{LL}-c^{(4d)}_{LL}]\\
\mu\frac{dc^{(2)}_{4L}}{d\mu}&=\frac1{6} N_cN_g\frac{g_2^2}{4\pi^2}c^{(5)}_{LL}\\
\mu\frac{dc^{(3)}_{4L}}{d\mu}&=-\frac1{9} N_cN_g\frac{g_1^2}{4\pi^2}[c^{(3)}_{LL}+2c^{(4u)}_{LL}-c^{(4d)}_{LL}]~, 
\end{aligned} 
\end{equation}
where we  have denoted by $N_c$ the number
of colors and by $N_g$ the number of quark generations. On the right
hand side of these equations we have
neglected terms proportional to $c^{(i)}_{4L}$ because we are considering a
situation with $c_{4L}(\LLFV)\ll c_{LL}(\LLFV)$. A full solution of the
RGE requires equations for $c^{(i)}_{LL}(\mu)$ too; however, since
$g_{1,2}$ are small, we retain only up to order $ g_{1,2}^2$  in the
amplitudes, and these terms drop out. 

In the second step, an effective theory with $U(1)$ gauge symmetry is
constructed with effective lagrangian given by matching at $\mu\sim
M_Z$ to the theory with $SU(2)\times U(1)$.  In this process $W^\pm$,
$Z^0$, and top quark are integrated out.  The relevant operators are
given in Eq.~(\ref{eq:ops-matched}), and the effective lagrangian in
Eq.~(\ref{eq:eff-lag-low}).  Matching coefficients gives 
\begin{equation}
\begin{aligned}
\label{eq:matching1}
\hat
c^{(3)}_{LL}(M_Z)&= c^{(3)}_{LL}(M_Z)- \frac16  
\sin^2\theta_w[c^{(1)}_{LL}(M_Z)+c^{(2)}_{LL}(M_Z)] \\
&  +\frac16\frac{\alpha}{\pi}Q_tF_tN_c[c^{(3)}_{LL}(M_Z)+
c^{(4u)}_{LL}(M_Z)-c^{(5)}_{LL}(M_Z)]\\
\hat
c^{(4u)}_{LL}(M_Z)&=c^{(4u)}_{LL}(M_Z)-
\frac23\sin^2\theta_w[c^{(1)}_{LL}(M_Z)+c^{(2)}_{LL}(M_Z)] \\
&  +\frac23\frac{\alpha}{\pi}Q_tF_tN_c[c^{(3)}_{LL}(M_Z)+
c^{(4u)}_{LL}(M_Z)-c^{(5)}_{LL}(M_Z)]\\
\hat
c^{(4d)}_{LL}(M_Z)&=c^{(4d)}_{LL}(M_Z)+
\frac13\sin^2\theta_w[c^{(1)}_{LL}(M_Z)+c^{(2)}_{LL}(M_Z)] \\
&  -
\frac13\frac{\alpha}{\pi}Q_tF_tN_c[c^{(3)}_{LL}(M_Z)+c^{(4u)}_{LL}(M_Z)-
c^{(5)}_{LL}(M_Z)]\\
\hat 
c^{(5)}_{LL}(M_Z)&=c^{(5)}_{LL}(M_Z)-
(\frac12-\frac12\sin^2\theta_w)[c^{(1)}_{LL}(M_Z)+c^{(2)}_{LL}(M_Z)]\\
&-\frac12\frac{\alpha}{\pi}Q_tF_tN_c[c^{(3)}_{LL}(M_Z)+
c^{(4u)}_{LL}(M_Z)-c^{(5)}_{LL}(M_Z)]\\
\hat
c^{(1)}_{4L}(M_Z)&=c^{(1)}_{4L}(M_Z)+c^{(2)}_{4L}(M_Z)-
(\frac12-\sin^2\theta_w)[c^{(1)}_{LL}(M_Z)+c^{(2)}_{LL}(M_Z)]\\
&-\frac{\alpha}{\pi}Q_tF_tN_c[c^{(3)}_{LL}(M_Z)+c^{(4u)}_{LL}(M_Z)-
c^{(5)}_{LL}(M_Z)]\\
\hat
c^{(3)}_{4L}(M_Z)&=c^{(3)}_{4L}(M_Z)+\sin^2\theta_w[c^{(1)}_{LL}(M_Z)+
c^{(2)}_{LL}(M_Z)]\\
&-\frac{\alpha}{\pi}Q_tF_tN_c[c^{(3)}_{LL}(M_Z)+c^{(4u)}_{LL}(M_Z)-
c^{(5)}_{LL}(M_Z)]\\
\hat c^{(1)}_{RL}(M_Z)&=c^{(1)}_{RL}(M_Z)-c^{(2)}_{RL}(M_Z)
\end{aligned} 
\end{equation}
%

Finally, for the third step we need the RGE satisfied by 
$\hat{c}^{(1)}_{4L}$  and $\hat{c}^{(3)}_{4L}$:
\begin{equation}
\mu\frac{d\hat c^{(1)}_{4L}}{d\mu}=\mu\frac{d\hat c^{(3)}_{4L}}{d\mu}=
-\frac{e^2N_c}{9\cdot4\pi^2}
 \left[(2n_u-n_d)\hat c^{(3)}_{LL}+2n_u\hat c^{(4u)}_{LL}-n_d\hat c^{(4d)}_{LL}-(2n_u+n_d)\hat c^{(5)}_{LL}\right]~.
\end{equation}
Here $n_u$ and $n_d$ again stand for the number of up-type and
down-type quarks, respectively. It is easy to check that if one adds
to the one loop amplitude from $\hat O^{(i)}_{LL}$, $i=3,\ldots,5$,
the tree level amplitude from $\hat O^{(1,3)}_{4L}$, and one does not
re-sum leading logs ({\it i.e.,} one works to order $e^2$ only), then
the $\mu$ dependence cancels in the amplitude (the $\mu$ dependence in
$\hat c^{(1,3)}_{4L}$ cancels that of the loop amplitude).

The solution to these equation is given in Sec.~\ref{sec-RG}, in
Eq.~(\ref{eq:wc4L}).

\section{A sum rule for $H_1^{r} (\mu_\chi,\mu)$}
\label{app:chpt}

Let us return  to the issue of matching the quark-level description
with the chiral effective theory calculation. In particular, in this
appendix we will show the explicit dependence of the chiral coupling
$H_1$ on the $\overline{\rm MS}$ renormalization scale $\mu$.

The light quark contribution to the matrix element $\langle 3e |
\hat{O}^{(i)}_{LL} | \mu \rangle$ in eq.~(\ref{eq:me1}) is UV
divergent, being proportional to $\Pi^{i,{\rm EM}}(q^2)$.  In
dimensional regularization the divergent correlator reads (work with
$i=4u$ for definiteness)
\begin{equation}
\Pi^{4u,{\rm EM}}(q^2) = \frac{N_c Q_u}{(4 \pi)^2} 
\Big[ \frac{4}{3} \left(\frac{2}{4 - d} - \gamma + \log{4 \pi} \right)
 - 8 F (m_u,m_u^2/q^2) \Big] \ , 
\label{eq:corrQCD} 
\end{equation}
with $F (m_u,m_u^2/q^2)$ defined in Eq.~(\ref{eq:Ffunction}).  
The $\overline{\rm MS}$ prescription amounts to subtracting the term
proportional to $\left(\frac{2}{4 - d} - \gamma + \log{4 \pi}
\right)$.  Writing $\Pi(q^2) = \Pi (q_0^2) + \Pi(q^2) - \Pi(q_0^2)$,
with $q_0^2$ in the regime of validity of perturbation theory, the
$\overline{\rm MS}$ subtracted correlator can be written as:
\begin{eqnarray}
\Pi^{4u,{\rm EM}}_{\overline{\rm MS}} (q^2) &=& 
- \frac{4}{3} \frac{N_c Q_u}{(4 \pi)^2} \left[ 
\log \frac{- 2 q_0^2}{\mu^2} - \frac{5}{3} \right] \ + \     
\Pi^{4u,{\rm EM}} (q^2) - \Pi^{4u,{\rm EM}} (q_0^2) \\
&=& 
- \frac{4}{3} \frac{N_c Q_u}{(4 \pi)^2} \left[ 
\log \frac{- 2 q_0^2}{\mu^2} - \frac{5}{3} \right] \ + \     
\hat{\Pi}^{4u,{\rm EM}} (q^2) - \hat{\Pi}^{4u,{\rm EM}} (q_0^2) 
\label{eq:corrMSbar2} 
\end{eqnarray}
where we have used the notation $\hat{\Pi}(q^2) = \Pi(q^2) - \Pi(0)$. 
The first term has the correct short distance $\mu$ dependence. 
The remaining terms can be expressed through 
a (once subtracted) dispersion relation as
\begin{equation}
\Pi^{4u,{\rm EM}} (q^2) - \Pi^{4u,{\rm EM}} (0) = 
\frac{q^2}{\pi} \int_{s_{\rm th}}^{\infty} ds 
\frac{ {\rm Im}  \Pi^{4u,{\rm EM}} (s)}{s ( s - q^2 - i \epsilon)}
\label{eq:disp1} 
\end{equation}
If $q^2 \ll \Lambda_{\chi SB}^2$, we can saturate the spectral
function with the $\pi \pi$ and $ K K $ threshold contributions, up to
higher order terms in $q^2/\Lambda_{\chi SB}^2$.  Doing so and
comparing with the chiral perturbation theory result in
eq.~(\ref{eq:orrchpt}) leads to a matching condition involving the low
energy constant $L_{10}$ and the contact term $H_1$:
\begin{equation}
\frac{8}{3} \left( 2 H_1^r (\mu_\chi,\mu) + L_{10}^r (\mu_\chi) \right) 
+\frac{2}{3 (4 \pi)^2} \log \left( \frac{m_K m_\pi}{\mu_\chi^2} \right) 
= 
\frac{4}{3} \frac{N_c Q_u}{(4 \pi)^2} \left[ 
\log \frac{- 2 q_0^2}{\mu^2} - \frac{5}{3} \right] \ 
+ \hat{\Pi}^{4u,{\rm EM}} (q_0^2) 
\label{eq:matching}
\end{equation}
A few comments are in order:
\begin{itemize}
\item 
The LHS does not depend on $\mu_\chi$ (due to chiral RG for $H_1$ and
$L_{10}$).  

\item The RHS does not depend on $q_0^2$. This can be seen by
separating the dispersion integral for $\hat{\Pi}^{4u,{\rm EM}}
(q_0^2)$ in an IR and UV region, and then using the free quark spectral
function in the UV regime. This gives back a logarithmic dependence on
$q_0^2$ that cancels the one in the other term.

\item
Eq.~(\ref{eq:matching}) explicitly shows how $H_{1}$ depends on the
short distance renormalization conventions and in particular on the
scale $\mu$.  It also displays a purely non-perturbative contribution 
to the matching, namely $\hat{\Pi}^{4u,{\rm EM}} (q_0^2)$. 
In absence of reliable esimates of $\hat{\Pi}^{4u,{\rm EM}} (q_0^2)$,  
in our numerics we use the experimental value for $L_{10}^r$ 
and a bound for $H_{1} (\mu_\chi=m_\rho,\mu= 1 \, {\rm GeV})$
based on naive dimensional analysis. 

\end{itemize}


\end{document}